\documentclass{aa}
\usepackage[utf8]{inputenc}
\usepackage[varg]{txfonts}
\bibpunct{(}{)}{;}{a}{}{,}

\newcommand{\disperse}{DisPerSE$\:$}
\begin{document}

\title{Characterising filaments in the SDSS volume from the galaxy distribution}
\author{Nicola Malavasi\inst{1}
\and
Nabila Aghanim\inst{1}
\and
Marian Douspis\inst{1}
\and
Hideki Tanimura\inst{1}
\and
Victor Bonjean\inst{1}}

\institute{Universit{\'e} Paris-Saclay, CNRS, Institut d'astrophysique spatiale, 91405, Orsay, France}

\date{Received: 3rd February 2020 / Accepted: 31st July 2020}

\abstract{Detecting the large-scale structure of the Universe based on the galaxy distribution and characterising its components is of fundamental importance in astrophysics but is also a difficult task to achieve. Wide-area spectroscopic redshift surveys are required to accurately measure galaxy positions in space that also need to cover large areas of the sky. It is also difficult to create algorithms that can extract cosmic web structures (e.g. filaments). Moreover, these detections will be affected by systematic uncertainties that stem from the characteristics of the survey used (e.g. its completeness and coverage) and from the unique properties of the specific method adopted to detect the cosmic web (i.e. the assumptions it relies on and the free parameters it may employ). For these reasons, the creation of new catalogues of cosmic web features on wide sky areas is important, as this allows users to have at their disposal a well-understood sample of structures whose systematic uncertainties have been thoroughly investigated. In this paper we present the filament catalogues created using the discrete persistent structure extractor (DisPerSE) tool in the Sloan Digital Sky Survey (SDSS), and we fully characterise them in terms of their dependence on the choice of parameters pertaining to the algorithm, and with respect to several systematic issues that may arise in the skeleton as a result of the properties of the galaxy distribution (such as Finger-of-God redshift distortions and defects of the density field that are due to the boundaries of the survey).}

\titlerunning{Filaments in the SDSS}
\authorrunning{Malavasi et al.}


\maketitle

\section{Introduction}
The cosmic web \citep{deLapparent1986,Bond1996} is a deeply interconnected network of structures that extends throughout the visible Universe. It is formed of galaxy clusters (which occupy node positions) that are linked by essentially one-dimensional structures called filaments (some low-mass groups possibly inhabit filaments; these groups are known as knots). Clusters are often grouped together inside more massive structures \citep[i.e. superclusters, see e.g.][]{Liivamaagi2012} that can also be found at the intersection of filamentary structures, but their definition is less univocal. Filaments are located at the intersections of planar structures called walls, which surround vast and empty regions denominated by voids.

While this large-scale structure of the Universe (LSS) is also composed of dark matter and gas, it was through the galaxy distribution that it has started to be detected. Galaxy clusters were the first cosmic web features to be identified and studied because they are easily detectable through various techniques. Only with the advent of wide-area spectroscopic redshift surveys have other structures such as filaments begun to be systematically identified. Surveys such as the Two-Degree Field Galaxy Redshift Survey \citep[2dFGRS,][]{Colless2001}, the Sloan Digital Sky Survey \citep[SDSS,][]{York2000}, the Galaxy And Mass Assembly survey \citep[GAMA,][]{Driver2009}, the Vimos Public Extragalactic Redshift Survey \citep[VIPERS,][]{Scodeggio2018}, or the COSMOS survey \citep{Scoville2007} have allowed us to obtain statistical samples of filaments and other LSS features. For example, \citet{Chen2016SDSS} and \citet{Tempel2014} have produced filament catalogues in the SDSS \citep[but see also the works by][some of which also used the same algorithm as we used here]{AragonCalvoPhD, Sousbie2011b, Rost2019, Shuntov2020photoweb, Kraljic2020}. Other works such as \citet{Kraljic2018GAMA} and \citet{Alpaslan2014} detected filaments in GAMA, while \citet{Malavasi2017} detected filaments in VIPERS. Additionally, \citet{Gott2005}, \citet{Iovino2016}, and \citet{Kraljic2018GAMA} also identified walls in the SDSS, COSMOS, and GAMA surveys, respectively, while several projects are devoted to the analysis of voids \citep[see e.g.][for a summary]{Colberg2008}. Recently, not only spectroscopic surveys, but also the increased precision of photometric redshifts \citep[e.g. in COSMOS and in the Canada-France-Hawaii Telescope Legacy Survey, CFHTLS,][]{Laigle2015, Coupon2009} allowed for the detection of filaments in volumes of the Universe up to $z \sim 1$ (\citealt{Laigle2018, Sarron2019}, but also \citealt{Darvish2014}).

The common element in all these works is the application of a cosmic web detection algorithm to a sample of galaxies. Detecting cosmic web features in the Universe based on the galaxy distribution essentially requires extracting topological information from the geometrical properties of the density field, which is sampled by using galaxies as tracers. A wide variety of methods exist to perform this task \citep[see e.g.][for a review and a comparison of some of them]{Libeskind2018}. Cosmic web detection algorithms implement different techniques to address common problems. In particular, the ultimate goal is to detect a set of structures that are anisotropic in shape (e.g. elongated filaments), multi-scale (filaments can extend from a few to 100 Mpc), and intricately connected \citep[see e.g.][]{Cautun2014}. \citet{Libeskind2018} classified the various methods into six groups based on their main characteristics: graph and percolation techniques (e.g. minimal spanning tree, MST, \citealt{Alpaslan2014}, T-ReX, \citealt{Bonnaire2019}), stochastic methods (e.g. BISOUS, \citealt{Tempel2014}), geometric Hessian-based methods (e.g. the multiscale morphology filter, MMF, \citealt{AragonCalvo2007}), scale-space Hessian-based methods (e.g. the already mentioned MMF and NEXUS+, \citealt{Cautun2013}), topological methods (e.g. the discrete persistent structure extractor, \disperse, \citealt{Sousbie2011a}), and phase-space methods (e.g. ORIGAMI, \citealt{Falck2012}).

In this paper we apply the \disperse cosmic web detection algorithm to the SDSS to obtain a catalogue of filaments in a representative volume of the Universe. We chose this method because it can detect structures in a topologically correct way (i.e. the detected filaments are actual topological entities that obey mathematical rules) directly from the galaxy distribution because it is (almost) parameter free and has an adaptive nature that responds well to the observational characteristics of the survey that is used in terms of sampling rate of tracer galaxies and uniformity of their distribution in redshift and on the plane of the sky (i.e. \disperse can work with sparse datasets or with distributions of galaxies in which the density of tracers changes dramatically within the sampled volume). The catalogues of filaments described here have been used in \citet{Malavasi2019,Tanimura2019disperse,Bonjean2019Filaments} and represent the first published general-purpose catalogues of filaments in the SDSS obtained with the \disperse method. We study the systematic uncertainties deriving from the \disperse method and the SDSS survey characteristics. We systematically characterise the effect of the \disperse parameters on the cosmic web reconstruction, and we fully describe the problems present in the catalogues and how to address them.

This paper is structured as follows: in Sect. \ref{data} we describe the SDSS survey and the galaxy samples that we used to detect the filaments, and in Sect. \ref{method} we summarise the \disperse$\:$ algorithm and its application. We describe the critical point and filament catalogues in Sect. \ref{catalogue_description} and we search for systematic effects due to the \disperse method. We validate the catalogue in Sect. \ref{validation} and present our conclusions and a summary in Sect. \ref{conclusions}. Throughout this paper we use a \citet{PlanckCollaborationXIIICosmoparams} cosmology, with $H_{0} = 67.74\: \mathrm{km}\: \mathrm{s}^{-1}\: \mathrm{Mpc}^{-1}$, $\Omega_{\mathrm{m}} = 0.3075$, and $\Omega_{\Lambda} = 0.6925$. Equatorial coordinates are given in the J2000 reference.

\section{Data}
\label{data}
When the best galaxy survey is chosen on which to run \disperse \ to create a filament sample, several factors come into play. The main characteristics that were sought after in the choice of the sample here were essentially the area coverage and the uniformity of the selection function. Among the available surveys, the Sloan Digital Sky Survey \citep[SDSS,][]{York2000} offered a wide area coverage and a large number of galaxies to exploit.

Although \disperse$\:$ can effectively handle inhomogeneous datasets and those with a non-contiguous coverage of the 3D volume of the Universe sampled, it is still recommended to work with galaxy samples that are as uniform as possible so as to provide an adequate sampling of the density field \citep[see e.g. the discussion of the treatment of gaps in the VIPERS survey region when \disperse\ was applied in][]{Malavasi2017}. For this reason, we focused on two galaxy samples extracted from the SDSS whose selection function was rather well understood: the main galaxy sample \citep[MGS,][]{Strauss2002} from the SDSS Data Release 7 (DR7) Legacy survey \citep{Abazajian2009}, and the LOWZ+CMASS sample \citep{Reid2016} from SDSS Data Release 12 (DR12, \citealt{Alam2015}).

Both these samples offer a rather uniform angular coverage on the plane of the sky and a good redshift sampling for the study of the 3D density distribution. Moreover, although the number density distribution as a function of redshift (hereafter $n(z)$) of the sources is not flat at all redshifts, it is quite stable in certain redshift ranges. We present the samples more in detail in the following sections.

\subsection{SDSS DR7 Legacy survey MGS}
The SDSS DR7 MGS \citep{Abazajian2009,Strauss2002} is a sample of $697\,920$ galaxies that are located mainly in the northern hemisphere, and a few stripes sample the southern hemisphere. Galaxies in this sample have both spectroscopic and photometric information. The sample selection described in \citet{Strauss2002} ensures that sources in the MGS have a Petrosian $r$-band magnitude $r_{\mathcal{P}} \le 17.77$ and an $r$-band Petrosian half-light surface brightness $\mu_{50} \le 24.5\, \mathrm{mag}\, \mathrm{arcsec}^{-2}$. The completeness of the sample is stated to be 99\%, while the error on the redshift measurement is lower than 30 km/s. The sample is publicly available on the SDSS website\footnote{\url{http://classic.sdss.org/legacy/index.html}}.

From this sample, we visually isolated the galaxies belonging to a contiguous region located in the northern hemisphere, so as to eliminate the isolated stripes in the southern hemisphere. We also selected sources with reliable redshifts in the form of \textsc{zwarning} = 0, \textsc{zconffinal} > 0.35, and \textsc{zfinal} > 0, following \citet{Strauss2002}. 

\begin{figure*}
\centering
\includegraphics[width = \linewidth, trim = 1cm 0cm 3cm 0cm, clip = true]{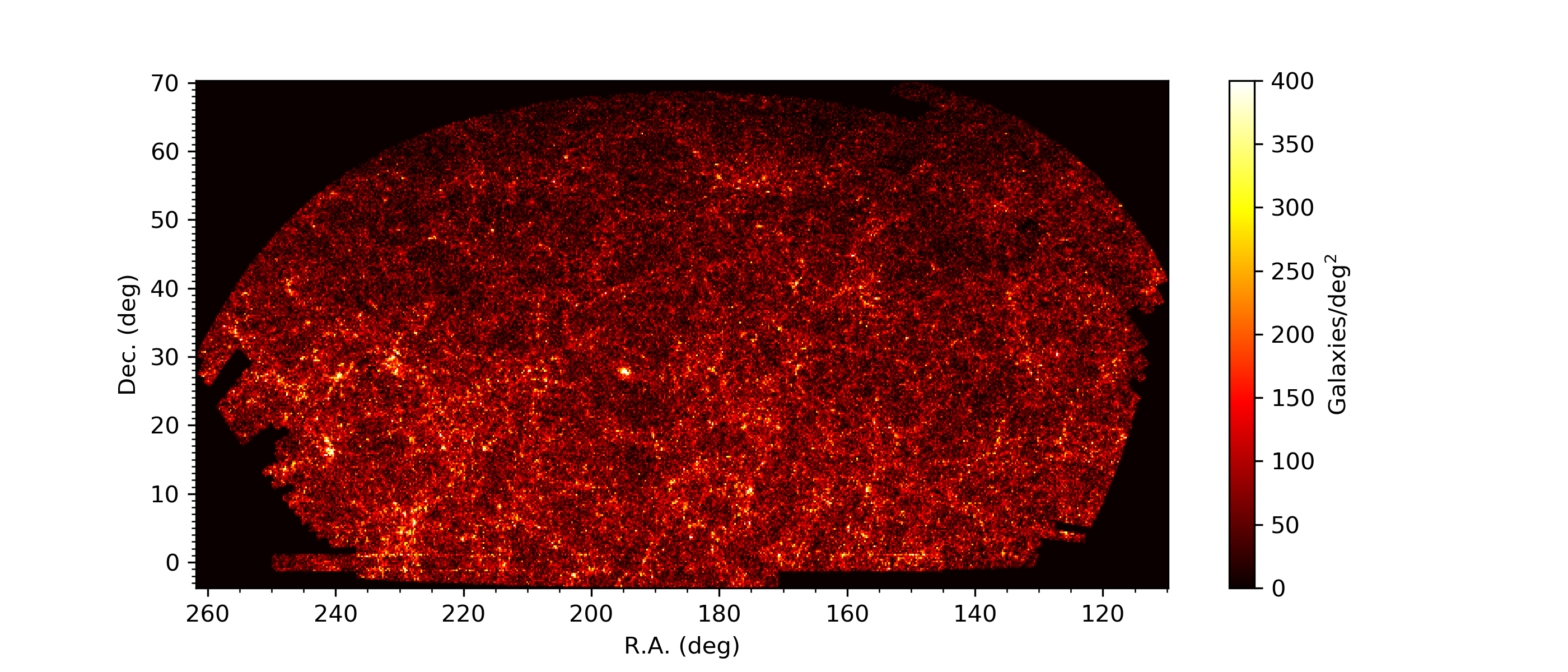}
\caption{Angular distribution of galaxies of the Legacy MGS in the selected part of the northern galactic hemisphere fed to \disperse. North is up, and east is to the left.}
\label{angdist_legacy}
\end{figure*}

The resulting angular distribution of the $566\,452$ sources constituting the sample that was input to the \disperse algorithm is shown in Figure \ref{angdist_legacy}. The distribution of the galaxies on the plane of the sky is rather uniform, with no clear holes or gaps. The imprint of the cosmic web on the galaxy distribution is already clearly visible, while the average surface density of sources of 78 $\mathrm{gal}/\deg^{2}$ is in line with the value of 92 $\mathrm{gal}/\deg^{2}$ reported by \citet{Strauss2002}.

\begin{figure}
\centering
\includegraphics[width = \linewidth, trim = 0cm 1cm 0cm 1cm, clip = true]{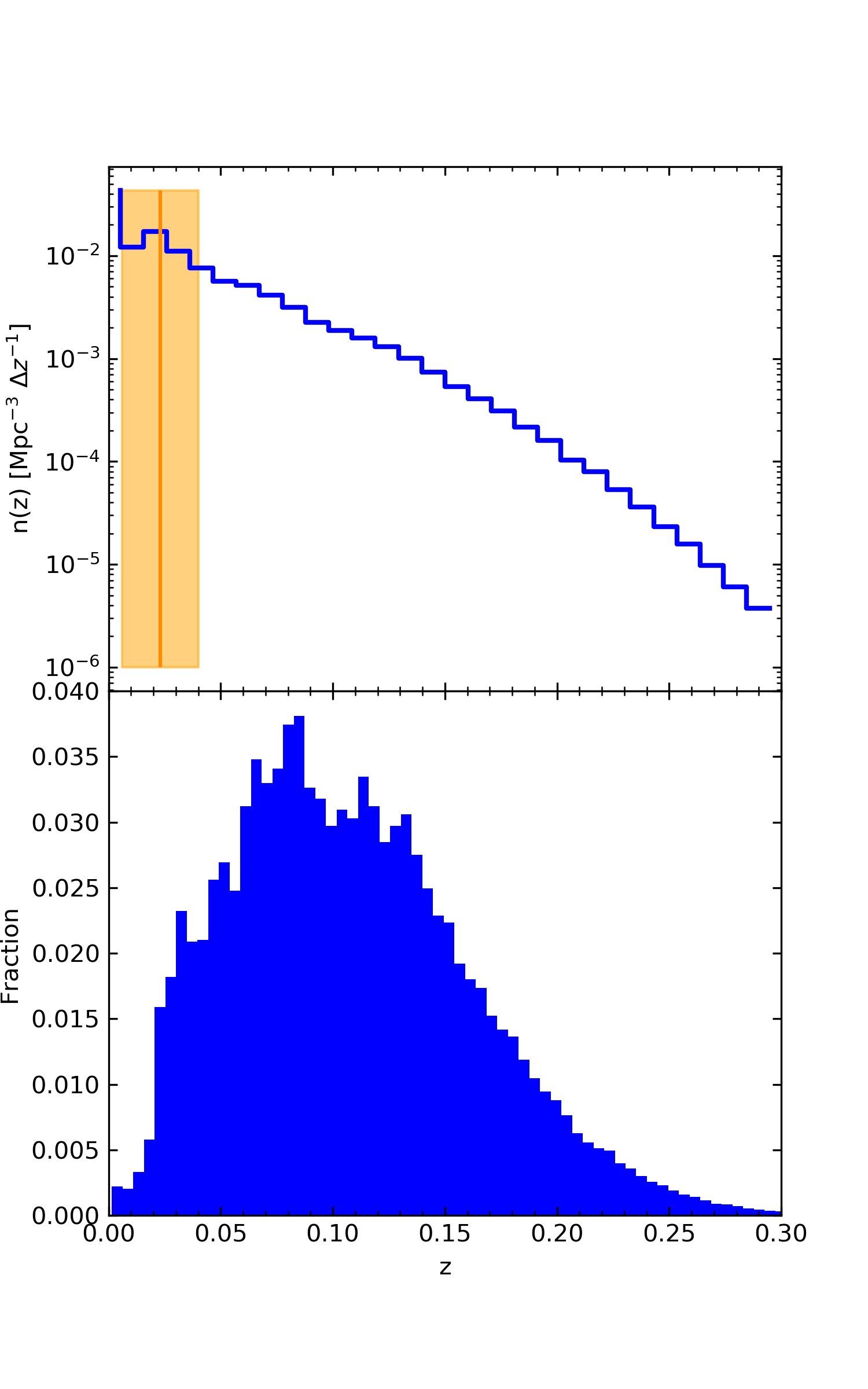}
\caption{Redshift distribution (bottom panel) and number density as a function of redshift in the survey volume ($n(z)$, top panel) for the MGS. The orange line and shaded area mark the region of interest for the analysis of the LSS around the Coma cluster of \citet{Malavasi2019}.}
\label{nofz_legacy}
\end{figure}

Figure \ref{nofz_legacy} shows the redshift distribution for the same galaxies (which ranges from $z = 0$ to $z \sim 0.3$, bottom panel) and the corresponding number density distribution in the survey volume. In order to compute it, we used the area relative to the northern galactic hemisphere elliptic footprint given in \citet{Tempel2014groups} of 7221 $\deg^{2}$. Although this value may not correspond to the actual effective area we used in this work, this only affects the normalisation of $n(z)$ and leaves the shape unchanged. This allowed us to verify that the number density of galaxies in the survey volume did not change significantly in the redshift range in which we carried out our analysis of the LSS around the Coma cluster ($z \in [0.006, 0.040]$, which is highlighted in the figure by the orange shaded region). The redshift distribution, on the other hand, confirms that the Coma analysis exploits a redshift range that is well covered by the MGS.

\begin{table}
\caption{Useful quantities for the galaxy samples drawn from the SDSS. The number of galaxies refers to the northern galactic hemisphere only for both surveys. The survey area was taken from the value for the northern hemisphere quoted in \citet{Tempel2014groups} (MGS) and from Table 2 of \citet[][effective area, NGC, CMASS survey]{Reid2016}. As there is no value for the combined LOWZ+CMASS area for the NGC but only the two separate values for LOWZ and CMASS, we chose the value for the area of CMASS as it is the largest of the two under the consideration that the LOWZ survey has a smaller area within the CMASS area.  The coordinates of the centre of the field ($\mathrm{R.A.}_{fc}, \mathrm{Dec.}_{fc}$) are used in the remaining paper to align the $x$-axis with the LoS direction when we convert from equatorial into Cartesian coordinates or vice versa.}
\label{useful_quantities}
\centering
\begin{tabular}{c c}
\hline\hline
Quantity & Value \\
\hline
\multicolumn{2}{c}{Legacy MGS} \\
\hline
Number of galaxies & $566\,452$ \\
Area ($\deg^{2}$) & 7221 \\
$\mathrm{R.A.}_{fc}, \mathrm{Dec.}_{fc} (\deg)$ & 186.183, 26.845 \\
Redshift range & $0 \div 0.3$ \\
\hline
\multicolumn{2}{c}{LOWZ+CMASS} \\
\hline
Number of galaxies & $953\,193$ \\
Area ($\deg^{2}$) & 6851 \\
$\mathrm{R.A.}_{fc}, \mathrm{Dec.}_{fc} (\deg)$ & 184.894, 28.153 \\
Redshift range & $0 \div 0.8$ \\
\end{tabular}
\end{table}

\subsection{SDSS DR12 LOWZ+CMASS sample}
We relied on the total LOWZ+CMASS sample described in \citet{Reid2016}\footnote{Freely available at \url{https://data.sdss.org/sas/dr12/boss/lss/}}, which is composed of $953\,193$ galaxies in the northern galactic hemisphere region and of $372\,542$ in the southern galactic hemisphere. These galaxies were selected through a series of cuts in the $(r-i)$ vs $(g-r)$ plane, and their model magnitudes (corrected for Milky Way extinction) are in the range $16 \le r_{cmod} \le 19.6$ (LOWZ) and $17.5 \le i_{cmod} \le 19.9$ (CMASS). We used only the galaxies in the northern galactic hemisphere here. 

\begin{figure*}
\centering
\includegraphics[width = \linewidth, trim = 1cm 0cm 3cm 0cm, clip = true]{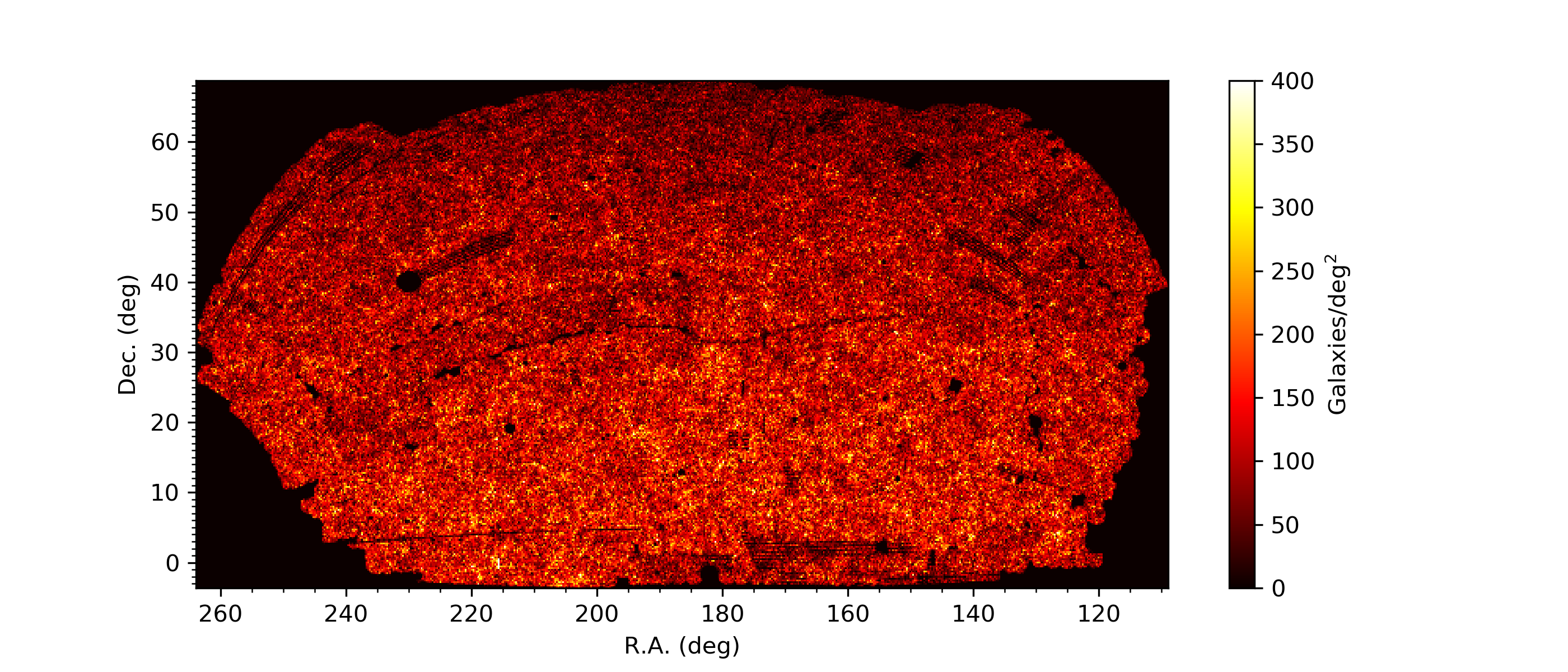}
\caption{Angular distribution of galaxies of the LOWZ+CMASS sample in the selected part of the northern galactic hemisphere fed to \disperse. North is up, and east is to the left.}
\label{angdist_lowzcmass}
\end{figure*}

The angular distribution of the galaxies in the northern galactic hemisphere is shown in Figure \ref{angdist_lowzcmass}. This sample provides a slightly less homogeneous coverage of the plane of the sky than the Legacy MGS, a few holes and empty stripes are visible. Still, the galaxy distribution provides a good sampling of the survey area, and the cosmic web imprint is again visible. The average surface density of galaxies is 139 $\mathrm{gal}/\deg^{2}$. This is in line with the value of 155 that is roughly estimated from the last line of Table 2 of \citet{Reid2016}, summing the number of targets$/\deg^{2}$ for the NGC for LOWZ and CMASS and considering that the two footprints occupy a similar area.

\begin{figure}
\centering
\includegraphics[width = \linewidth, trim = 0cm 1cm 0cm 1cm, clip = true]{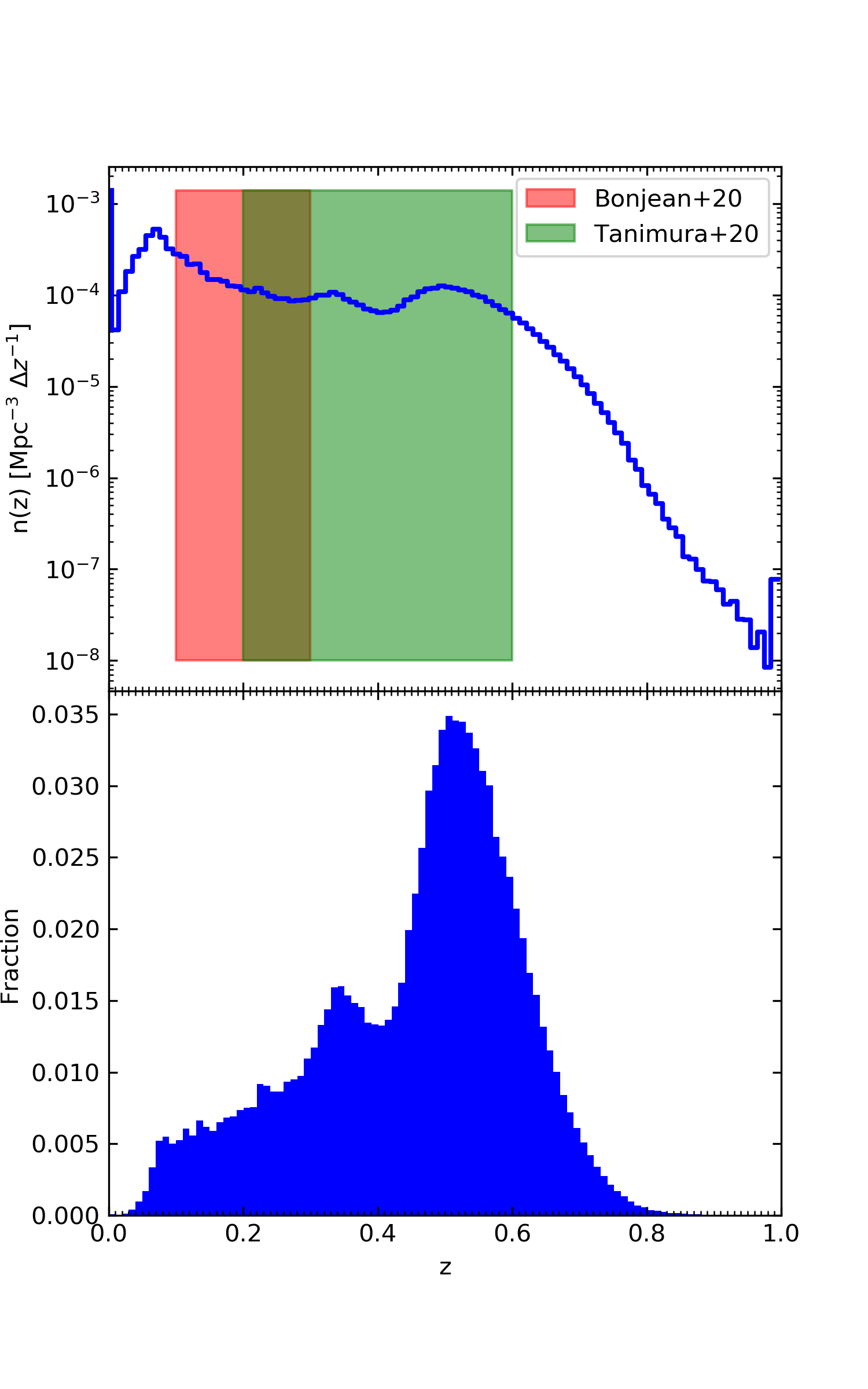}
\caption{Redshift distribution (bottom panel) and number density as a function of redshift in the survey volume ($n(z)$, top panel) for the LOWZ+CMASS sample. The shaded areas mark the redshift ranges explored by  \citet[][red,]{Bonjean2019Filaments} and \citet[][green]{Tanimura2019disperse}.}
\label{nofz_lowzcmass}
\end{figure}

Spectroscopic information is available for all the objects in the sample, and the redshift distribution and number density of sources in the survey volume are shown in Figure \ref{nofz_lowzcmass}. The redshift distribution of the LOWZ+CMASS sample extends to higher redshifts than the MGS, ranging from $z = 0$ to $z \sim 0.8$. The completeness is stated to be $99\%$ for CMASS and $97\%$ for LOWZ \citep[see Figure 8 of][]{Reid2016}. The number density in the survey volume (top panel of Figure \ref{nofz_lowzcmass}) was computed using an area of 6851 $\deg^{2}$ \citep[Table 2 of][]{Reid2016}, considering that the LOWZ survey is contained within the CMASS survey. Again, choosing an incorrect value for the area has the effect of slightly changing the normalisation of $n(z)$ in Figure \ref{nofz_lowzcmass} but leaves the shape unaltered. After an initial peak, the number density in the survey volume is rather flat, especially in the redshift range covered by \citet{Bonjean2019Filaments} ($z \in [0.1,0.3]$) and \citet{Tanimura2019disperse} ($z \in (0.2,0.6)$).

The samples were converted from equatorial into Cartesian coordinates prior to applying the \disperse algorithm. The conversion was made while also applying a rotation of the system so that the $x$-axis points in the line-of-sight (LoS) direction towards the centre of the field. The coordinates for the centre of the field were computed as the average of the galaxy coordinates and are listed in Table \ref{useful_quantities} together with other quantities for the samples. This transformation was applied to all figures that show galaxies or filament distributions in the remaining paper.

\subsection{Cluster samples}
\label{subsectionclusters}
When we validated the filament catalogue, we relied on several samples of galaxy clusters that were detected at various wavelengths and are described in Section 2 of \citet{Malavasi2019}. We briefly summarise their properties here. The number of clusters in each sample and their redshift range are given in Table \ref{cluster_quantities}.

\begin{table}
\caption{Number of sources and redshift range covered for the various cluster and group samples.}
\label{cluster_quantities}
\centering
\begin{tabular}{c c c}
\hline\hline
Sample & Number of clusters & Redshift range \\
\hline
MXCX & 1743 & $0.003 \div 1.3$ \\
Planck SZ & 1745 & $0.01 \div 1.7$ \\
\citet{Tempel2017} & 6873 & $0.002 \div 0.2$ \\
redMaPPer & $15\,694$ & $0.06 \div 0.9$ \\
\citet{Wen2012} & $95\,684$ & $0.02 \div 0.8$ \\
\citet{Wen2015} & $25\,419$ & $0.05 \div 0.75$ \\
\end{tabular}
\end{table}

We relied on clusters detected in the X-ray that are assembled in the Meta-Catalogue of X-ray-detected Clusters of galaxies \citep[MCXC,][]{Piffaretti2011}. This is a homogenised compilation of 1743 clusters in the redshift range $z \sim 0.003 \div 1.3$ and covers the full sky, constructed from pre-existing cluster catalogues based on serendipitous observations and on the ROSAT All-Sky Survey \citep{Voges1999}.

In the microwave domain, we relied on the publicly available Sunyaev-Zel'dovich (SZ) cluster database\footnote{\url{http://szcluster-db.ias.u-psud.fr}}. This catalogue is the union of several catalogues of clusters that were detected through their  SZ signal with several instruments: {\it Planck} \citep[the early, first, and second catalogues of SZ sources, ESZ, PSZ1, PSZ2,][]{PlanckCollaborationVIIIESZ, PlanckCollaborationXXIXClusters, PlanckCollaborationXXVIIClusters}, the South Pole Telescope \citep[SPT,][]{Williamson2011,Reichardt2013,Ruel2014,Bleem2015} the Atacama Cosmology Telescope \citep[ACT,][]{Hasselfield2013}, the Arcminute Microkelvin Imager \citep[AMI,][]{AMIConsortium2012, AMIConsortium2013, AMIConsortium2013bis, Schammel2013}, and the Combined Array for Research in Millimeter-wave Astronomy \citep[CARMA,][]{Brodwin2015, Buddendiek2015}. The catalogue is composed of 1745 galaxy clusters with a confirmed redshift and measured physical parameters, distributed over the full extent of the sky.

We also relied on samples of optical clusters and groups, constructed from galaxy catalogues of the SDSS survey. In particular, \citet{Tempel2017} ran a friend-of-friends algorithm on the galaxy distribution from the SDSS Data Release 12 and identified groups and clusters in the redshift range $z \sim 0.002 \div 0.2$. They identified $88\,662$ groups and clusters ($68\,887$ have a number of galaxies members $N_{gal} \gtrsim 3$). We eliminated from the sample all clusters with fewer than $N_{gal} = 6$ galaxies, resulting in 6873 remaining objects. The richness threshold was chosen following \citet{Tempel2017} in order to select only clusters with a reliable mass estimate. The resulting mass range is $10^{11} \div 10^{15} M_{\sun}$. Using data from the SDSS Data Release 8 combined with the redMaPPer algorithm, \citet{Rykoff2014} detected $26\,111$ clusters, of which we used $15\,694$ that have a measured redshift (covering the range $z \sim 0.06 \div 0.9$). Also based on SDSS Data Release 8 and Data Release 12, \citet{Wen2012} and \citet{Wen2015} identified $132\,684$ and $25\,419$ groups and clusters (of which $95\,684$ have a measured redshift in the case of the \citealt{Wen2012} sample). These objects cover the redshift range $z \sim 0.02 \div 0.8$ and $z \sim 0.05 \div 0.75$, respectively.

\section{Method}
\label{method}

\subsection{Detecting the filaments: DisPerSE}
To construct our filament catalogues, we used the DisPerSe method. The \disperse algorithm is capable of operating in three dimensions on discrete, non-smoothed datasets. The detection of filaments and other elements of the cosmic web relies on a two-step process. First the galaxy density field is measured, then the discrete Morse theory is applied and filaments are detected.

We measured the density field of the galaxy distribution using the Delaunay tessellation field estimator \citep[DTFE,][]{SchaapWeygaert2000, WeygaertSchaap2009}. This method uses Delaunay tessellation to cover the space with tetrahedrons, using the galaxy positions as vertexes. The density field can be recursively smoothed by averaging the value at each galaxy position with the density field values of the galaxies that are directly connected to it through an edge of the tessellation. We refer to a single iteration of the smoothing process as a smoothing cycle. The smoothing process can be iterated several times by averaging the already-averaged density values. In the figures, we refer to no smoothing of the density field, one smoothing cycle, and two smoothing cycles as SD0, SD1, and SD2, respectively. The \disperse method then proceeds with the application of the discrete Morse theory to the measured density field. While details of the algorithm can be found in \citet{Sousbie2011a} and \citet{Sousbie2011b}, we summarise the main features here. 

The first step consists of computing the gradient of the density field and of identifying points at which the gradient vanishes (critical points). They can be classified into maxima of the density field, minima, and type 1 and type 2 saddles (local density minima bounded to structures, such as walls or filaments, respectively). Filaments are defined as field lines of constant gradient connecting critical points (maxima and saddles). Special types of critical points called bifurcation points were inserted at the position where filaments intersected in post-processing. The final result is a set of critical points connected by filaments that are composed of short segments of which the positions of the extrema are given. The length of the segments is related to the typical length of the edges of the tessellation.

Another strong point of the \disperse method is the application of the persistence theory to eliminate spurious filaments that are likely to be due to the Poisson noise of the discrete galaxy distribution. Critical points are coupled in topological constructs called persistence pairs, based on the ratio of their density as measured by the DTFE. Persistence pairs are eliminated from the real data set when they are closer to the noise persistence distribution than a certain number of $\sigma$. This procedure is equivalent to imposing a signal-to-noise ratio threshold for the filaments. In the following we compare two different persistence cuts (a low-persistence and a high-persistence cut) for our filaments, that is, for a $3\sigma$ and a $5\sigma$ cut\footnote{Except for the LOWZ+CMASS filaments, for which one smoothing cycle is applied to the density field before DisPerSE is applied. In this case, a $5\sigma$ threshold results in a crash of the algorithm because it tries to cancel too many low-persistence pairs. We set the high-persistence cut to $4.5\sigma$ in this case.}. We would like to point out that what we define as our low-persistence cut ($3\sigma$) is still high enough to eliminate most of the spurious filaments (see e.g. the discussion in \citealt{Sousbie2011a} and \citealt{Malavasi2017}, who stated that  the probability that a feature is spurious is $\sim 5\%$ at the $2\sigma$ level and drops to $0.006\%$ at $4\sigma$). We applied the high-persistence cut to ensure that our results hold also when only the most secure filaments are selected.

The \disperse algorithm allows for a spatial smoothing of the skeleton after its detection. This procedure was carried out in a similar way as the smoothing of the density field by averaging the spatial positions of the extrema of the segments that compose the filaments, each segment position averaged with the positions of the segments directly attached to it. The procedure can be iterated by averaging the already-averaged positions as done for the density field.

\subsection{Edge effects}
\label{method_edgeeffects}
The edges of a survey footprint present a problem when an environmental estimation has to be made. The abrupt lack of tracers beyond a certain limit in space means that the density field is underestimated close to the borders, and any analysis using this quantity will be affected by this, including topological reconstructions of the LSS. The \disperse algorithm has several ways to handle boundary conditions, depending on the sample it is applied to. In particular, when the algorithm is run, it is possible to choose smooth boundary conditions. Additional particles are added outside of the initial domain, with a random distribution that follows the interpolated density field at the survey edge. These so-called guard particles are used to compute the tessellation outside the initial bounding box. \disperse then tags all the features (critical points and the filaments connected to them) as belonging to the boundary in which one feature of the tessellation belongs to the guard particles or whose value of the DTFE density may be affected by the connecting of the tesselation with particles outside of the domain \citep[see Section 7 of][]{Sousbie2011a}. As bifurcation points are added in post-processing after the skeleton has been extracted from the data, they cannot be tagged by \disperse even if they are on the boundary.

\begin{figure}
\centering
\includegraphics[width = \linewidth]{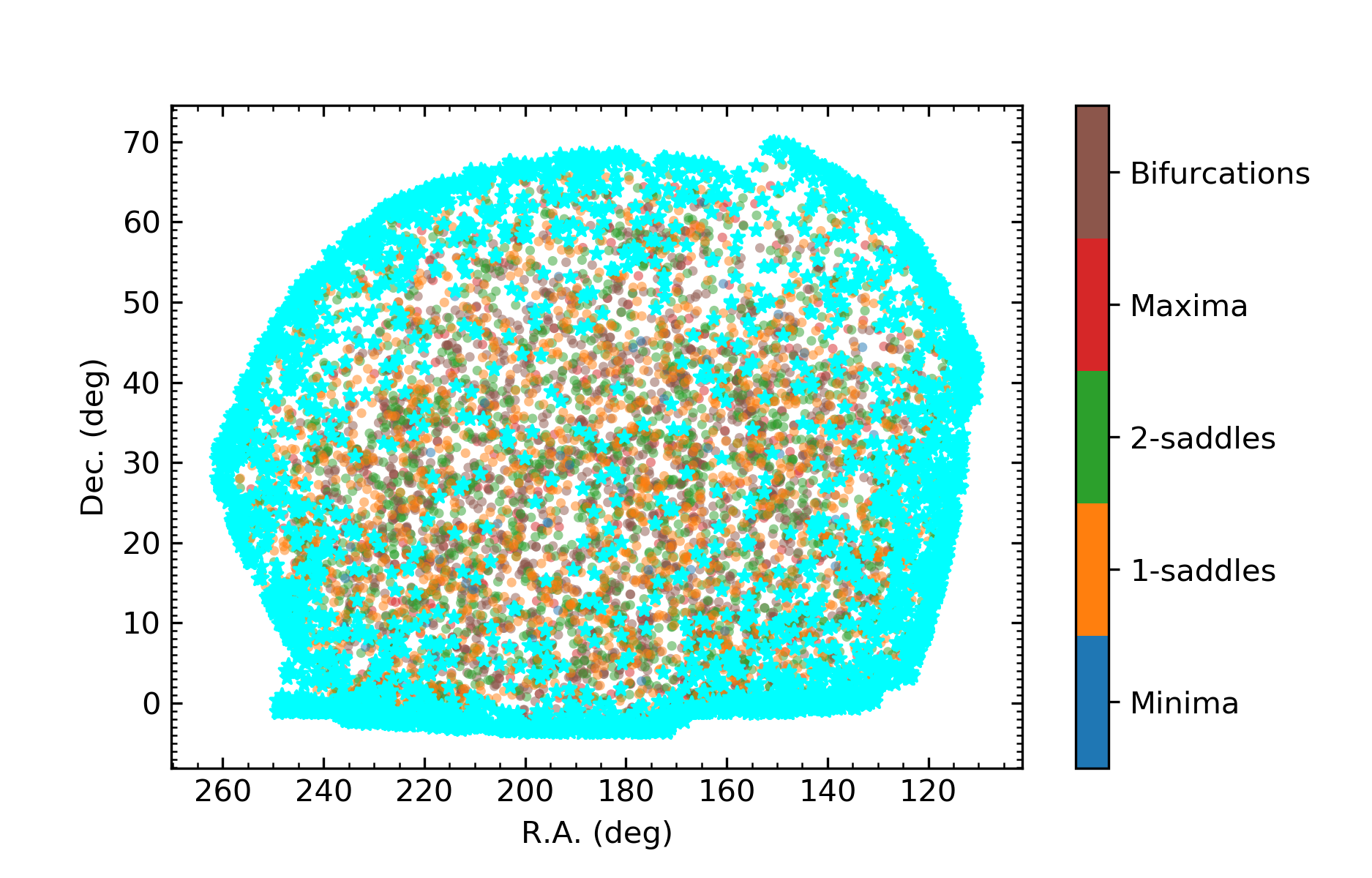}
\includegraphics[width = \linewidth]{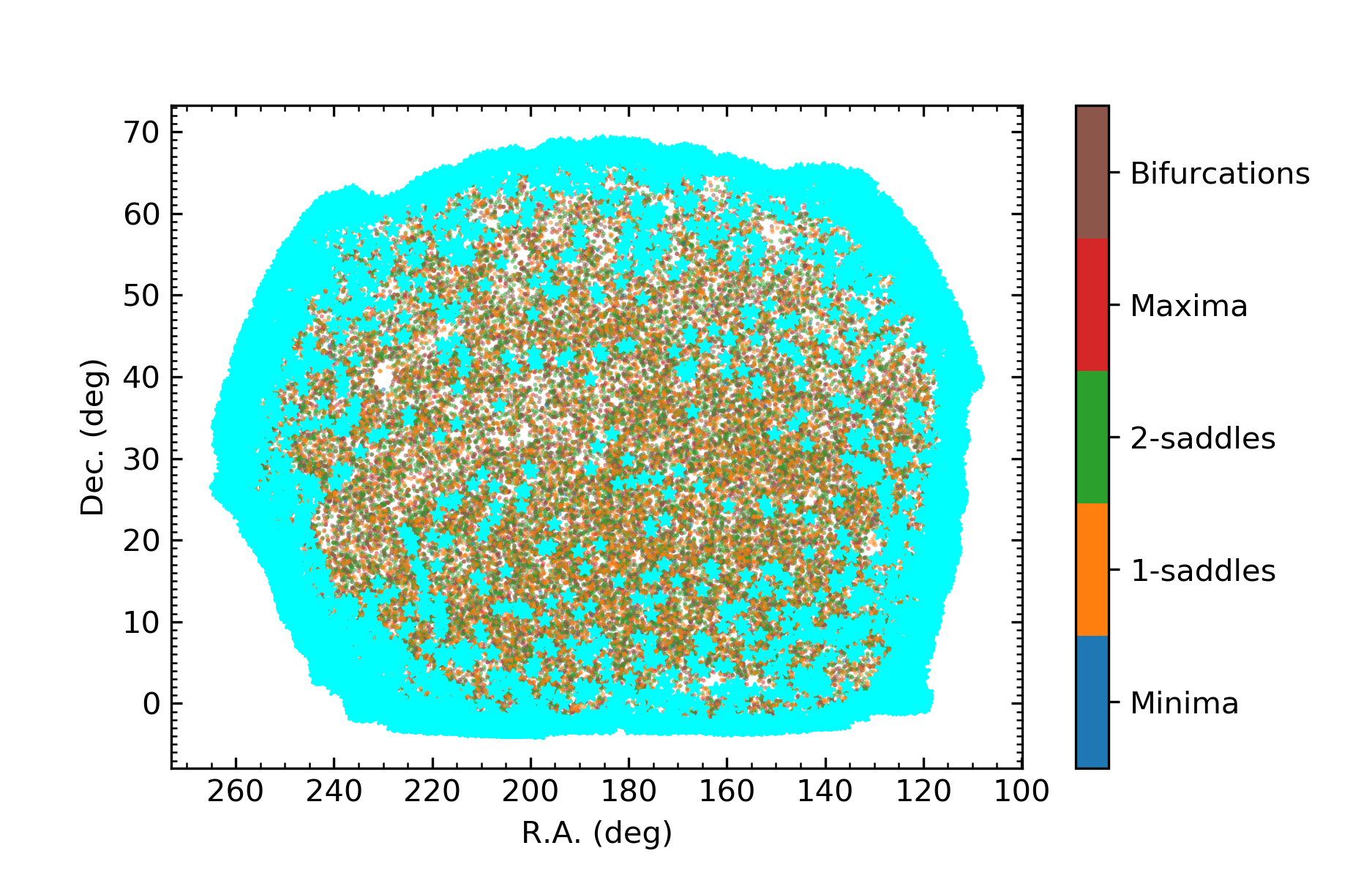}
\caption{Angular distribution on the plane of the sky of critical points that are affected by edges. The top panel refers to the Legacy MGS and the bottom panel to LOWZ+CMASS. In each panel all the critical points detected by \disperse are shown with filled circles, colour-coded according to their type. Cyan stars highlight critical points that are affected by edges (including bifurcations, see text). As an example, the $3\sigma$ persistence skeletons have been chosen, with one smoothing cycle of the density field in the Legacy MGS case and no smoothing at all of the density field in the LOWZ+CMASS case.}
\label{whatremoved_boundary}
\end{figure}

We eliminated from the sample of critical points all those that are tagged as boundary by \disperse,$\:$ as well as all the bifurcation points that are within 100 Mpc (200 Mpc) from the surface that defines the convex hull of all the critical points tagged as boundary for the Legacy MGS (LOWZ+CMASS). This is a very conservative cut, but it allows us to determine the effect of boundaries on our skeleton reconstruction. Figure \ref{whatremoved_boundary} shows the critical points (including bifurcations) that were removed from the sample as belonging to the boundaries in the full area of the Legacy MGS and of the LOWZ+CMASS sample. Filaments connected to these critical points were removed consistently. In the bottom panel of this figure, critical points do not seem to be tagged as boundary more often close to the large holes which are visible within the footprint of the galaxy distribution, for example in Figures \ref{angdist_lowzcmass} and \ref{Fil_map_radec_lowzcmass} (see below). This is due to the DTFE method that we used to measure the density field: as stated also in Section \ref{fil_cat_sec}, the tetrahedrons of the tessellation are able to cross gaps and holes, which essentially means that they interpolate the density field over regions that are devoid of tracers. For this reason, these regions are not considered as boundary. 

\begin{figure}
\centering
\includegraphics[width = \linewidth]{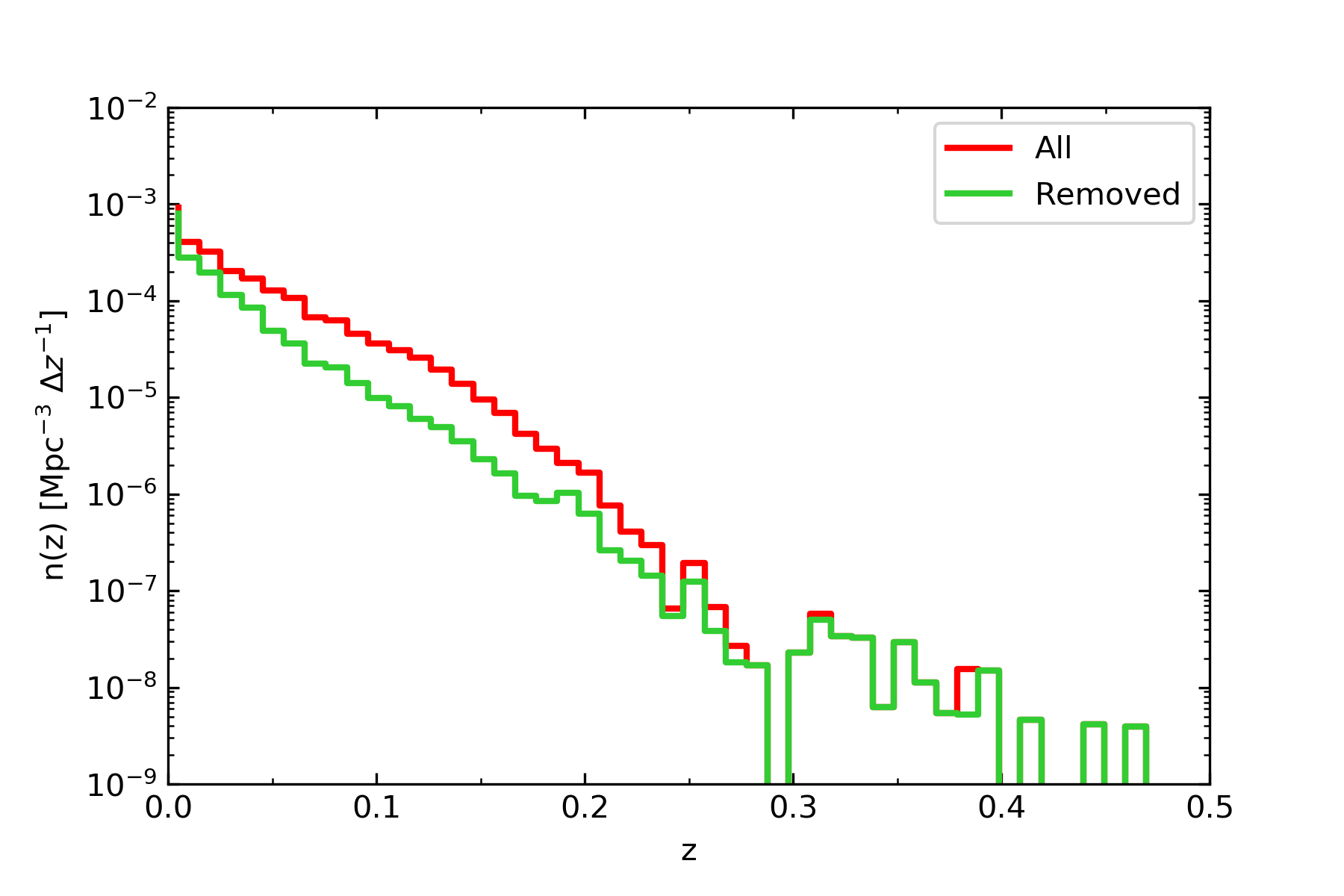}
\includegraphics[width = \linewidth]{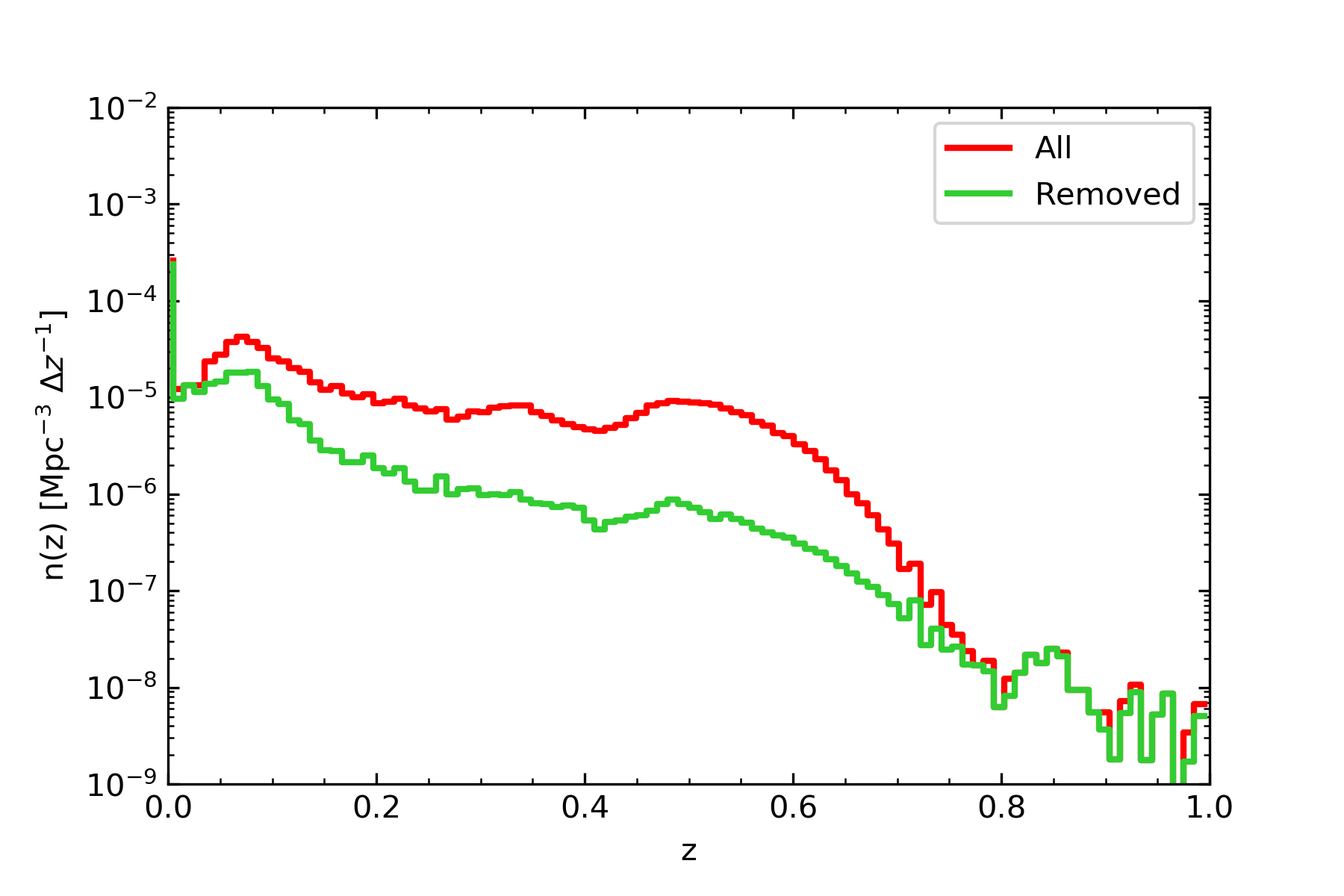}
\caption{Number density distribution as a function of redshift of all critical points and of those affected by edges. The top panel refers to the Legacy MGS and the bottom panel to LOWZ+CMASS. In each panel the number density distribution as a function of redshift of critical points that are affected by edges (including bifurcations, see text) is marked in green. The number density distribution as a function of redshift of all critical points in the sample, regardless of their type, is marked in red. As an example, the $3\sigma$ persistence skeletons have been chosen, with one smoothing cycle of the density field in the Legacy MGS case and no smoothing at all of the density field in the LOWZ+CMASS case.}
\label{whatremoved_boundary_zdist}
\end{figure}

Figure \ref{whatremoved_boundary_zdist} shows the number density distribution as a function of redshift of the critical points (including bifurcations) that were removed as belonging to the boundaries. This is compared to the total number density distribution as a function of redshift of all critical points in the skeleton. Both for the Legacy MGS and the LOWZ+CMASS samples, the $n(z)$ distribution of the removed critical points covers the full redshift range, and the shape is similar to that of the total $n(z)$ distribution. The absence of any particular feature in the number density distribution as a function of redshift of the removed critical points indicates that they were eliminated uniformly at all redshifts.

The numbers of critical points removed from the samples with this criterion (including bifurcations) are reported in Table \ref{cpremovededges}. We would like to stress that while for other types of critical points, \disperse provides a tag to determine whether they are affected by borders, bifurcations are added in post-processing and lack this identification tag. The distance threshold that we adopted here of 100 Mpc (200 Mpc) from the surface which defines the convex hull of all the critical points tagged as boundary, is an arbitrary choice. The sole intent in doing so was showing the effect (or lack thereof) of edge-affected points on the skeleton properties. This fixed threshold may even remove too many bifurcations, especially at low redshift, and a redshift-evolving criterion to remove bifurcations may be more indicated for analyses that involve this sample.

\begin{table}
\caption{Number of critical points that was removed from the samples because of edge effects. In parentheses we report the fractions of critical points that we removed from each skeleton. Although indicated as $5\sigma$, the high-persistence cut for LOWZ+CMASS in the one-smoothing case has been limited to $4.5\sigma$. The numbers shown here are purely indicative of the tests that we performed in order to validate the catalogue, and they represent a very conservative threshold.}
\label{cpremovededges}
\centering
\begin{tabular}{c c c}
\hline\hline
Smoothing & $3\sigma$ & $5\sigma$ \\
\hline
\multicolumn{3}{c}{Legacy MGS} \\
\hline
No smoothing & 10452 (22\%) & 4340 (45\%) \\
one smoothing & 3532 (35\%) & 1483 (68\%) \\
two smoothings & 1591 (52\%) & 579 (90\%) \\
\hline
\multicolumn{3}{c}{LOWZ+CMASS} \\
\hline
No smoothing & 8750 (13\%) & 3414 (44\%) \\
one smoothing & 2911 (26\%) & 1547 (65\%) \\
two smoothings & 1308 (45\%) & 517 (88\%)  \\
\hline
\end{tabular}
\end{table}

\subsection{Minor problems}
\label{method_minorissues}
We performed other checks for minor systematic problems that may affect the skeleton. These include minor issues that may not be important when analyses of statistical properties of the cosmic web are performed, but they may nevertheless affect the general quality of our reconstruction.

\begin{figure}
\centering
\includegraphics[width = \linewidth]{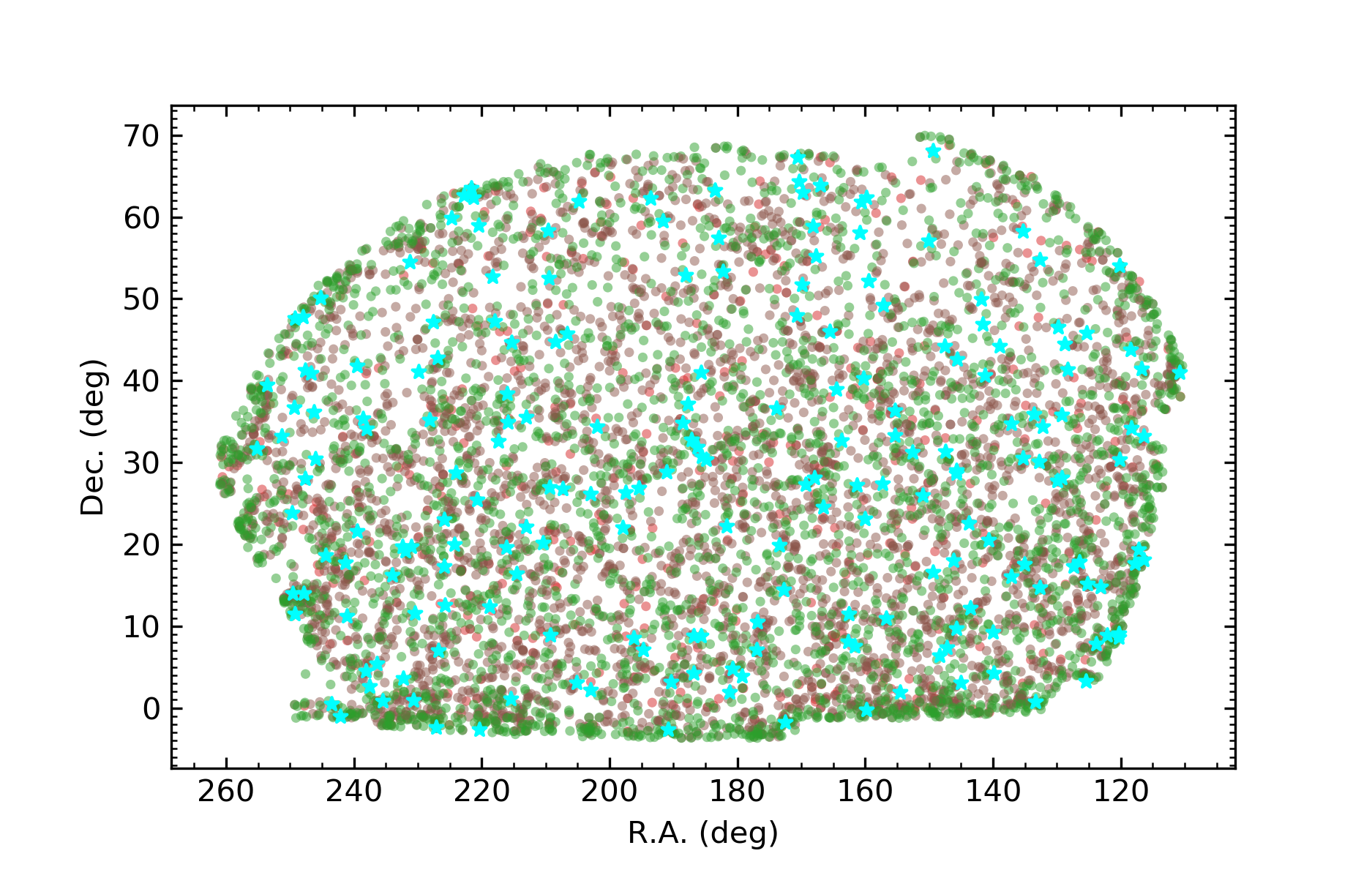}
\includegraphics[width = \linewidth]{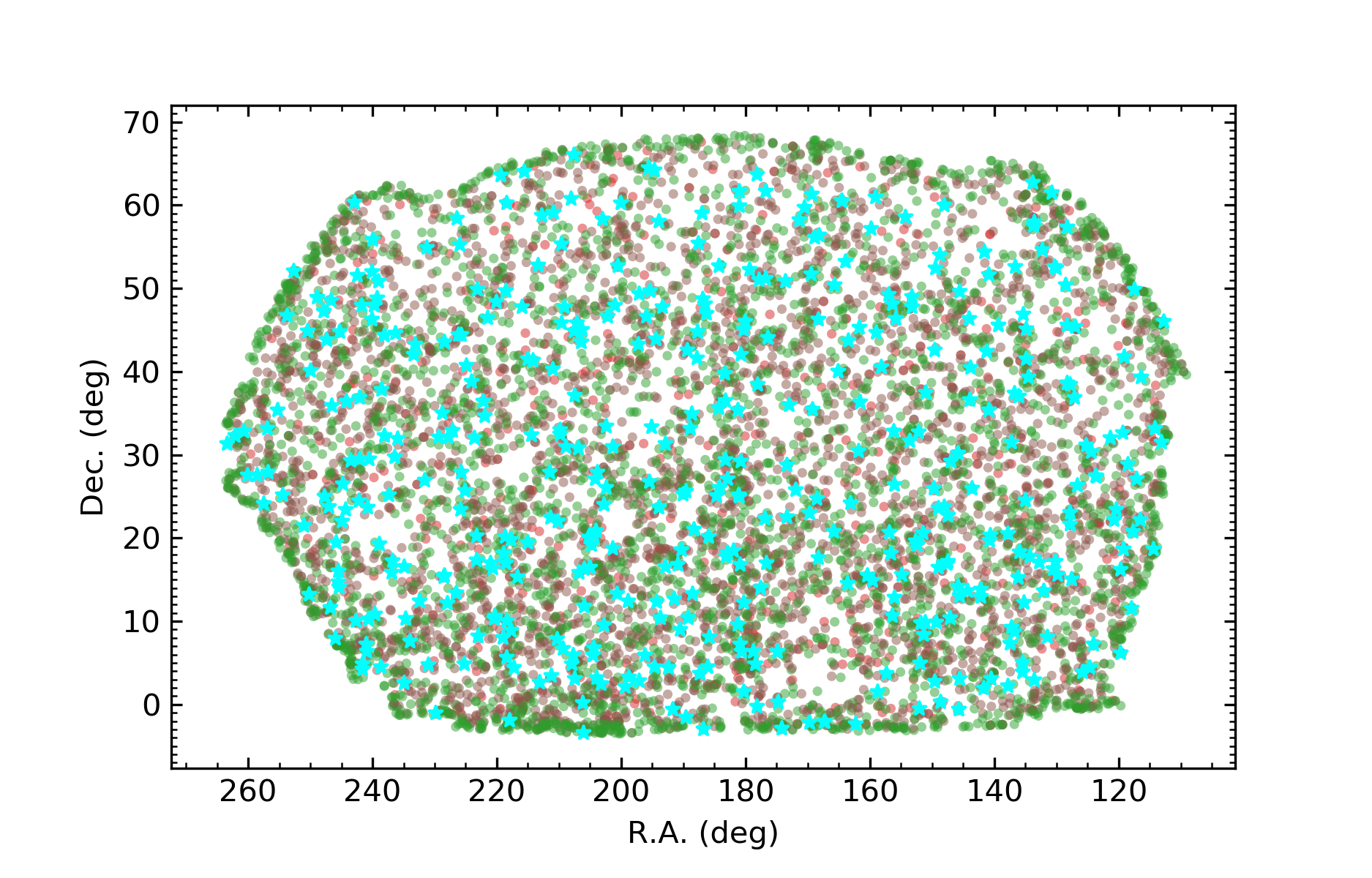}
\caption{Angular distribution on the plane of the sky of critical points affected by minor problems. The top panel refers to the Legacy MGS and the bottom panel to LOWZ+CMASS. In each panel, critical points detected by \disperse are shown with filled circles, colour-coded according to their type (green: type 2 saddles, red: maxima, brown: bifurcations). Only points of type maximum, type 2 saddle, or bifurcation are shown here as minima, and type 1 saddle points are not affected by the minor problems we consider here. Cyan stars highlight critical points that are affected by minor problems (i.e. superposing maxima and saddles that are connected by a filament of 0 Mpc or maxima connected to only one filament, see text). As an example, the $3\sigma$ persistence skeletons have been chosen, with one smoothing cycle of the density field for both the Legacy MGS and LOWZ+CMASS.}
\label{cleanup_example}
\end{figure}

We began by eliminating saddles and maxima from the skeleton that overlapped perfectly and were connected by a filament of exactly 0 Mpc. These topological defects are likely caused by the Poisson noise of the galaxy distribution, which creates sharp changes in the density field that the \disperse algorithm has trouble interpreting. Clearly, 0 Mpc filaments between overlapping critical points are non-physical, and we removed them from the skeleton. When we performed this cleaning of the skeleton, we eliminated the saddle and the 0-length filaments, which left only the maximum that is reconnected to other existing filaments.

We also eliminated isolated maxima that were connected by only one filament. These points are less problematic than overlapping saddles and maxima, as they most likely represent local maxima that are connected to the rest of the skeleton. These isolated nodes do not fully represent our expectations for the cosmic web, however, where clusters are not the end point of a single filament where matter accumulates, but rather intersection points of two or more filaments that channel matter accretion onto the structure. As \disperse is supposed to reconstruct a fully connected skeleton, these isolated maxima could present a problem (but a very minor one), and it is worth investigating how they affect our reconstruction of the cosmic web.

\begin{table}
\caption{Number of critical points and filaments that we removed from the samples because of minor problems. Although indicated as $5\sigma$, the high-persistence cut for LOWZ+CMASS in the one-smoothing case has been limited to $4.5\sigma$.}
\label{numbersremovedminor}
\centering
\begin{tabular}{c c c}
\hline\hline
Smoothing & $3\sigma$ & $5\sigma$ \\
\hline
\multicolumn{3}{c}{Legacy MGS} \\
\hline
No smoothing & 1833 & 108 \\
one smoothing & 206 & 6 \\
two smoothings & 66 & 1 \\
\hline
\multicolumn{3}{c}{LOWZ+CMASS} \\
\hline
No smoothing & 3290 & 161 \\
one smoothing & 450 & 22 \\
two smoothings & 146 & 3  \\
\hline
\end{tabular}
\end{table}

The cleaning of the skeleton described above results in the numbers, reported in Table \ref{numbersremovedminor}, of critical points and filaments that were removed from the Legacy MGS sample and the LOWZ+CMASS sample. These numbers do not include the critical points that might be affected by boundary effects as they were introduced in Section \ref{method_edgeeffects}. Figure \ref{cleanup_example} shows an example of the clean-up procedure on the Legacy MGS and LOWZ+CMASS skeletons. The critical points we eliminated are randomly distributed on the plane of the sky, indicating that these problems are not due exclusively to the edges.

\begin{figure}
\centering
\includegraphics[width = \linewidth]{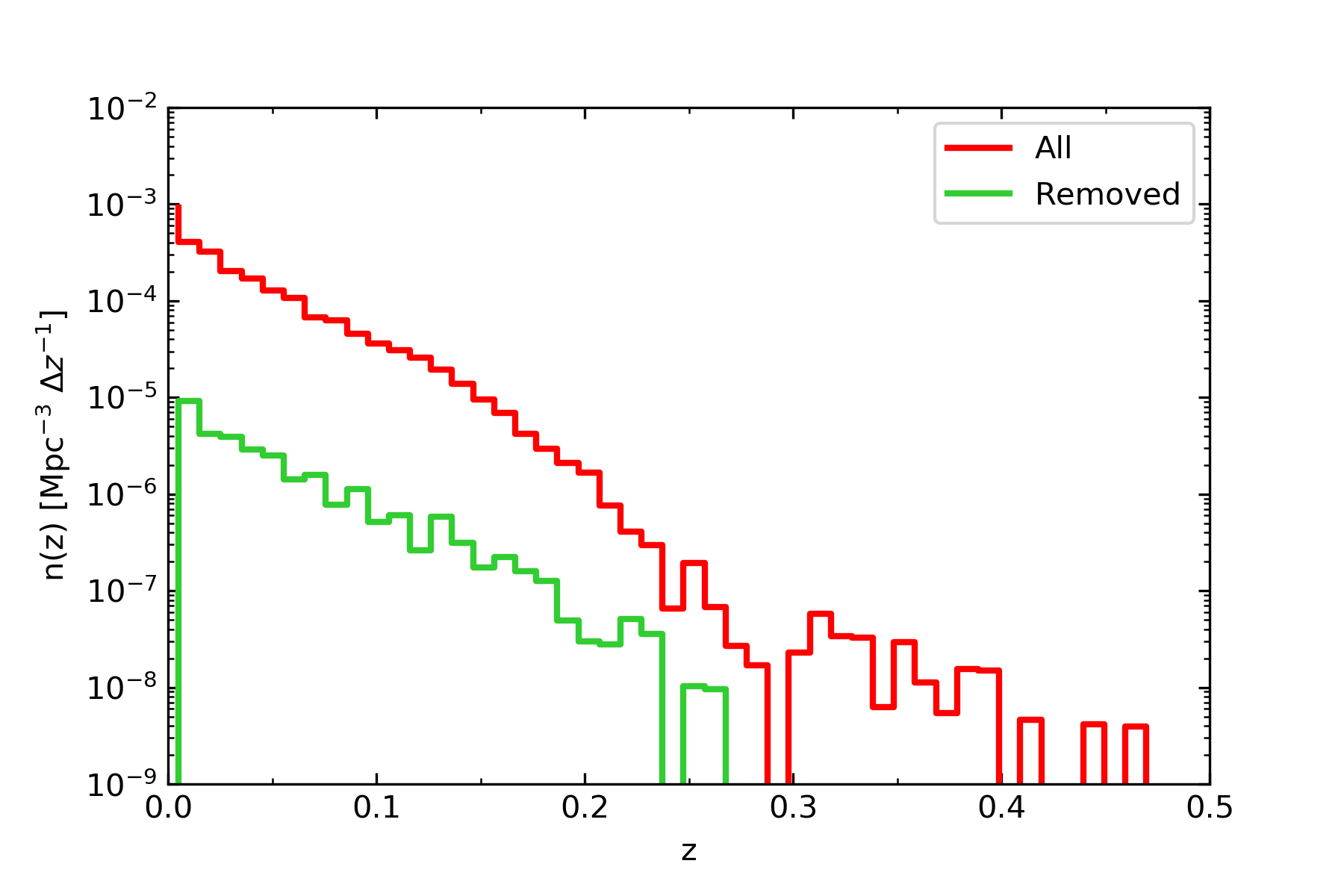}
\includegraphics[width = \linewidth]{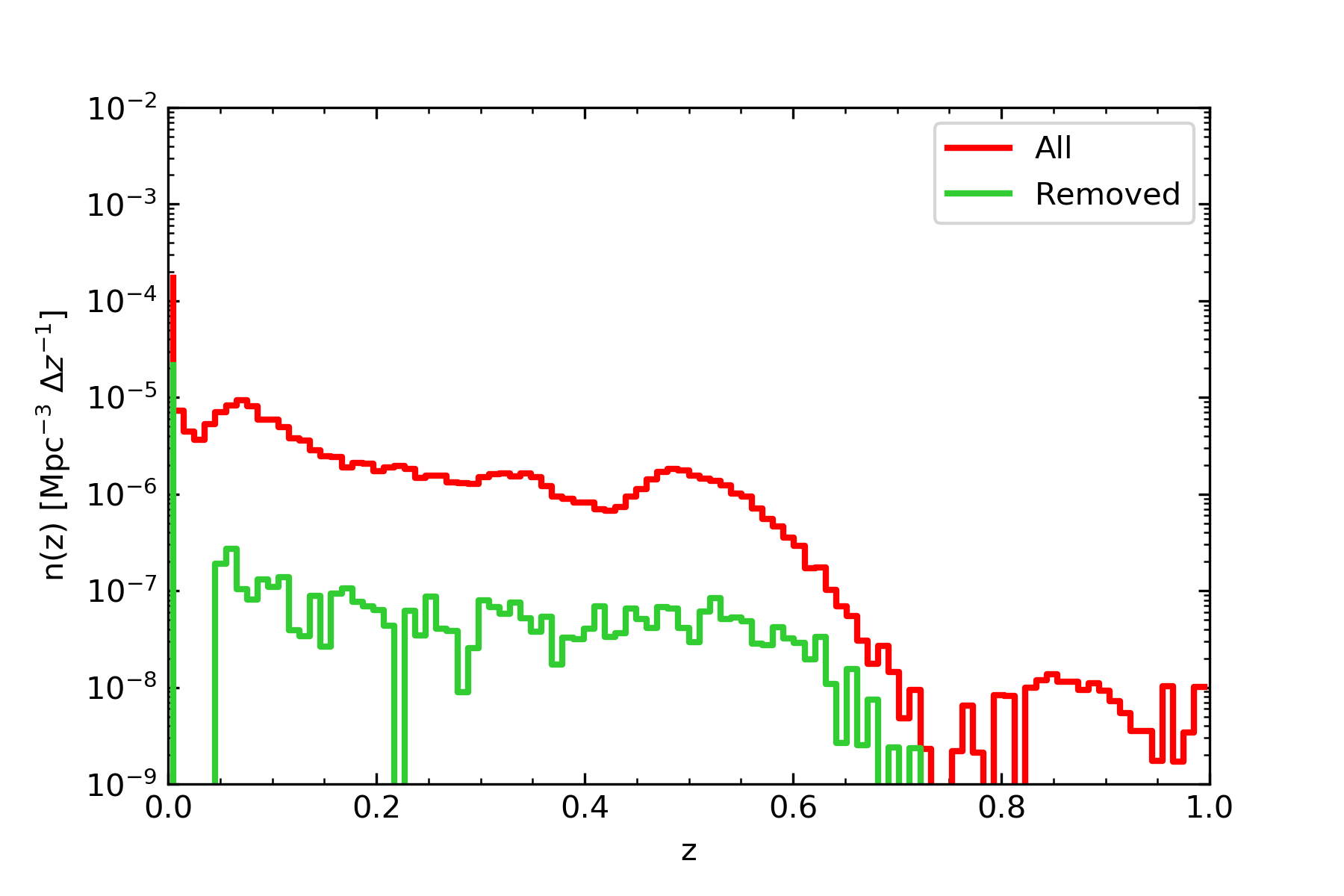}
\caption{Number density distribution as a function of redshift of all critical points and of those affected by minor problems. The top panel refers to the Legacy MGS and the bottom panel to LOWZ+CMASS. In each panel the number density distribution as a function of redshift of the critical points that are affected by minor problems (see text) is marked in green. The number density distribution as a function of redshift of all critical points in the sample, regardless of their type, is marked in red. As an example, the $3\sigma$ persistence skeletons with one smoothing cycle of the density field have been chosen.}
\label{whatremoved_minor_zdist}
\end{figure}

We report the number density distribution as a function of redshift of the critical points we eliminated from the sample compared to the number density distribution as a function of redshift of all critical points in Fig. \ref{whatremoved_minor_zdist}. The $n(z)$ distributions of the eliminated critical points and of the total sample are very similar, showing that when the sample is cleaned, critical points are eliminated uniformly across the redshift range.

\section{Description of the catalogue}
\label{catalogue_description}
The LSS catalogue we constructed is composed of a few thousand to a few tens of thousand critical points, depending on the number of smoothing cycles and the persistence (see Table \ref{ncrits}), as well as of the filaments that connect them (Table \ref{nfils}). Filaments connect maxima of the density field (which can be identified with galaxy clusters) to type 2 saddle points (i.e. local density minima located on filaments) or bifurcations (i.e. points where several filaments intersect, which can be identified with low-mass clusters and groups, which are often unresolved). Type 1 saddle points are local density minima located on walls that surround low-density regions centred on minima of the density field (which can be identified with the centres of voids). While in the analyses described in \citet{Malavasi2019}, \citet{Tanimura2019disperse}, and \citet{Bonjean2019Filaments} we only focused on maxima, type 2 saddle points, and bifurcations and in the present catalog we only distribute filaments and disregard walls for the moment, type 1 saddle points, and minima (voids), these types of critical points are still included in the numbers given in Table \ref{ncrits} and are included in some of the distributions presented below to provide a complete overview of the properties of the catalogue.

\begin{table}
\caption{Number of critical points in the samples, broken down by critical point type. Although indicated as $5\sigma$, the high-persistence cut for LOWZ+CMASS in the one-smoothing case has been limited to $4.5\sigma$.}
\label{ncrits}
\centering
\begin{tabular}{c c c c}
\hline\hline
Type & No smoothing & one smoothing & two smoothings \\
\hline
\multicolumn{4}{c}{Legacy MGS critical points} \\
\hline
\multicolumn{4}{c}{$3\sigma$}\\
\hline
All & 46667 & 10179 & 3081 \\
Minima & 337 & 264 & 183 \\
Type 1 saddles & 9554 & 2603 & 859 \\
Type 2 saddles & 15523 & 3012 & 875 \\
Maxima & 6286 & 663 & 194 \\
Bifurcations & 14967 & 3637 & 970 \\
\hline
\multicolumn{4}{c}{$5\sigma$}\\
\hline
All & 9647 & 2185 & 641 \\
Minima & 108 & 96 & 65 \\
Type 1 saddles & 2672 & 698 & 234 \\
Type 2 saddles & 2750 & 610 & 166 \\
Maxima & 184 & 11 & 1 \\
Bifurcations & 3933 & 770 & 175 \\
\hline
\multicolumn{4}{c}{LOWZ+CMASS critical points} \\
\hline
\multicolumn{4}{c}{$3\sigma$}\\
\hline
All & 66388 & 10986 & 2918 \\
Minima & 447 & 220 & 133 \\
Type 1 saddles & 12818 & 2559 & 709 \\
Type 2 saddles & 22154 & 3353 & 904 \\
Maxima & 9811 & 1003 & 317 \\
Bifurcations & 21158 & 3851 & 855 \\
\hline
\multicolumn{4}{c}{$5\sigma$}\\
\hline
All & 7685 & 2367 & 590 \\
Minima & 99 & 80 & 44 \\
Type 1 saddles & 2117 & 740 & 202 \\
Type 2 saddles & 2263 & 698 & 171 \\
Maxima & 236 & 29 & 4 \\
Bifurcations & 2970 & 820 & 169 \\
\end{tabular}
\end{table}

\begin{table}
\caption{Number of filaments in the samples. Although indicated as $5\sigma$, the high-persistence cut for LOWZ+CMASS in the one-smoothing case has been limited to $4.5\sigma$.}
\label{nfils}
\centering
\begin{tabular}{c c c c}
\hline\hline
Smoothing & $3\sigma$ & $5\sigma$ \\
\hline
\multicolumn{3}{c}{Legacy MGS filaments} \\
\hline
No smoothing & 44198 & 8353 \\
one smoothing & 8959 & 1590 \\
two smoothings & 2363 & 355 \\
\hline
\multicolumn{3}{c}{LOWZ+CMASS filaments} \\
\hline
No smoothing & 63391 & 6376 \\
one smoothing & 9764 & 1705 \\
two smoothings & 2288 & 351 \\
\end{tabular}
\end{table}

\subsection{Critical points}
As expected, the number of critical points decreases with the adopted persistence threshold, as less significant points are removed from the sample, and also with the smoothing. Introducing a smoothing of the density field reduces the amount of noise, and small-scale fluctuations are removed, together with their corresponding critical points.

\begin{figure}
\centering
\includegraphics[width = \linewidth]{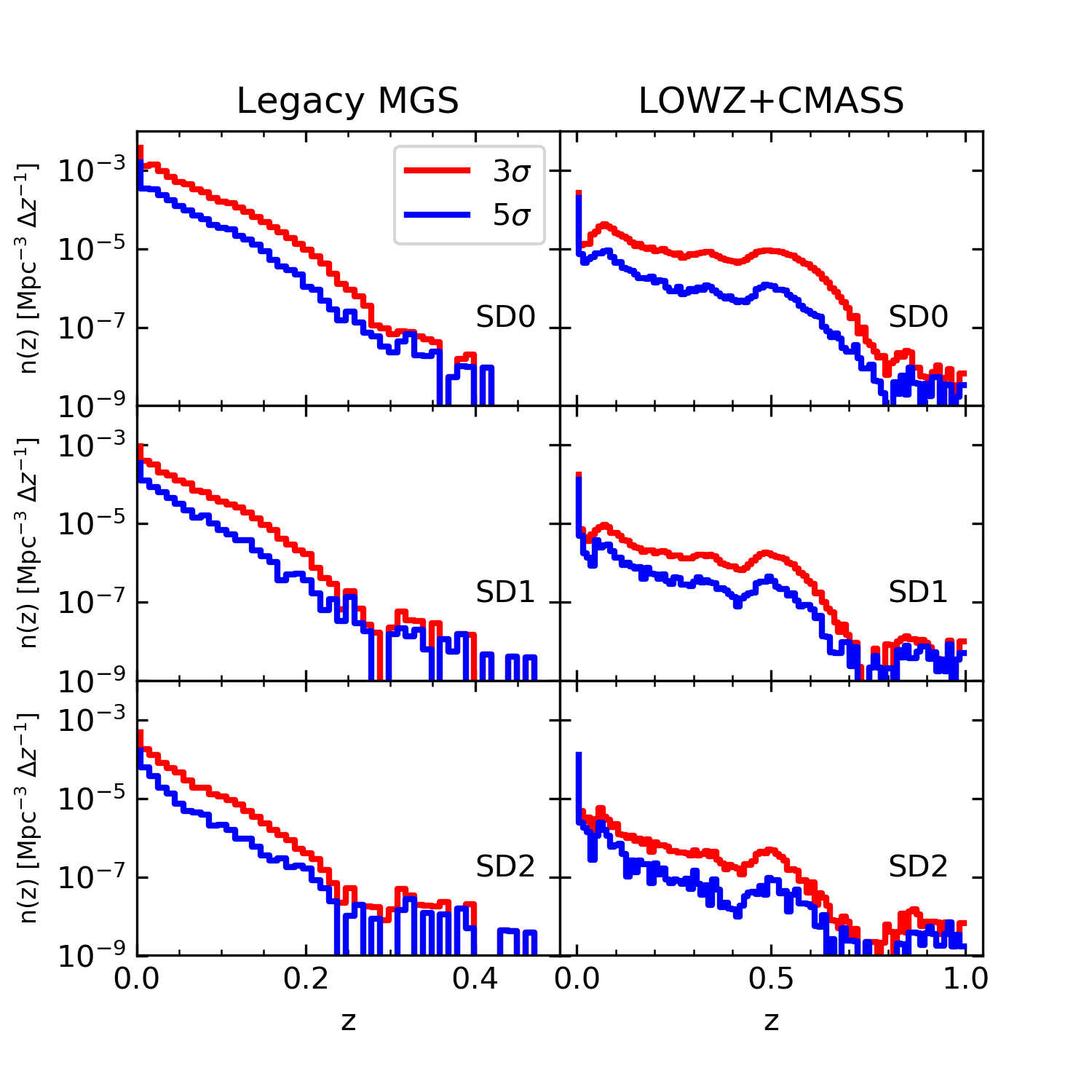}
\caption{Number density distribution as a function of redshift of the critical points. The left and right columns refer to the Legacy MGS and LOWZ+CMASS sample, respectively. Rows are different smoothing cycles of the density field prior to filament detection. Red lines refer to a $3\sigma$ persistence threshold, and blue lines refer to a $5\sigma$ persistence threshold. The explored range in redshift in the left and right column is different. Although indicated as $5\sigma$, the high-persistence cut for LOWZ+CMASS in the one-smoothing case has been limited to $4.5\sigma$.}
\label{redshift_dist}
\end{figure}

As expected, the number density distribution as a function of redshift of the critical points (Fig. \ref{redshift_dist}) closely follows the galaxy distribution because critical points provide a topological description of the density field traced by the galaxies. Nevertheless, changing the smoothing or the persistence threshold does not alter the shape of the $n(z)$ distribution of the critical points, indicating that critical points are removed uniformly in space. The only difference is in a lower density of critical points (i.e. in the normalisation of the curves) as increasing the persistence threshold indeed removes points from the sample.

\begin{figure}
\centering
\includegraphics[width = \linewidth]{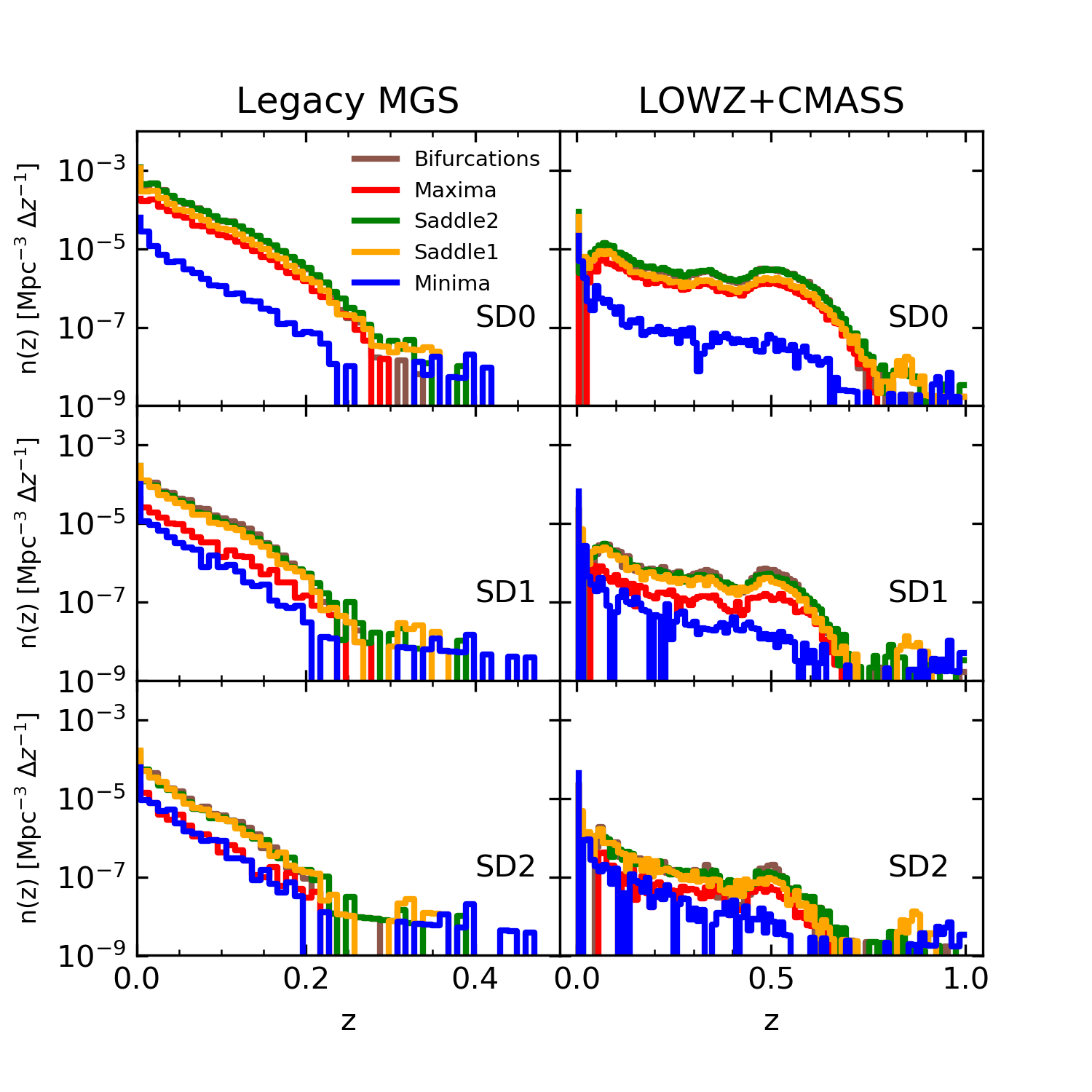}
\caption{Number density distribution as a function of redshift of the different types of critical points. The left and right columns refer to the Legacy MGS and LOWZ+CMASS sample, respectively. Rows are different smoothing cycles of the density field prior to filament detection. In each panel, the solid lines refer to maxima (red), bifurcations (brown), type 2 saddles (green), type 1 saddles (yellow), and minima (blue). Only the $3\sigma$ persistence threshold is shown here for clarity.}
\label{redshift_dist_type}
\end{figure}

Figure \ref{redshift_dist_type} shows the number density distribution as a function of redshift of the critical points when they are classified according to their type. Only the $3\sigma$ persistence threshold is shown here for clarity, as Fig. \ref{redshift_dist} showed that increasing the persistence threshold does not alter the shape of the number density distributions significantly. In the Legacy MGS case, all critical points follow the same $n(z)$ distribution, regardless of their type. This shows that the \disperse algorithm is able to consistently identify the various types of features of the density field throughout the full redshift range of the survey. In the case of the LOWZ+CMASS sample, this is also true when no smoothing is applied to the density field prior to the critical point extraction. Progressively applying an increasing degree of smoothing to the density field reduces the capability of \disperse of recovering low-density features (this problem is most evident in the case of minima) at high redshift. For both samples, minima are consistently the less numerous type of critical points detected by \disperse in the survey volume.

\begin{figure*}
\centering
\includegraphics[width = \linewidth, trim = 0cm 2cm 0cm 2cm, clip = true]{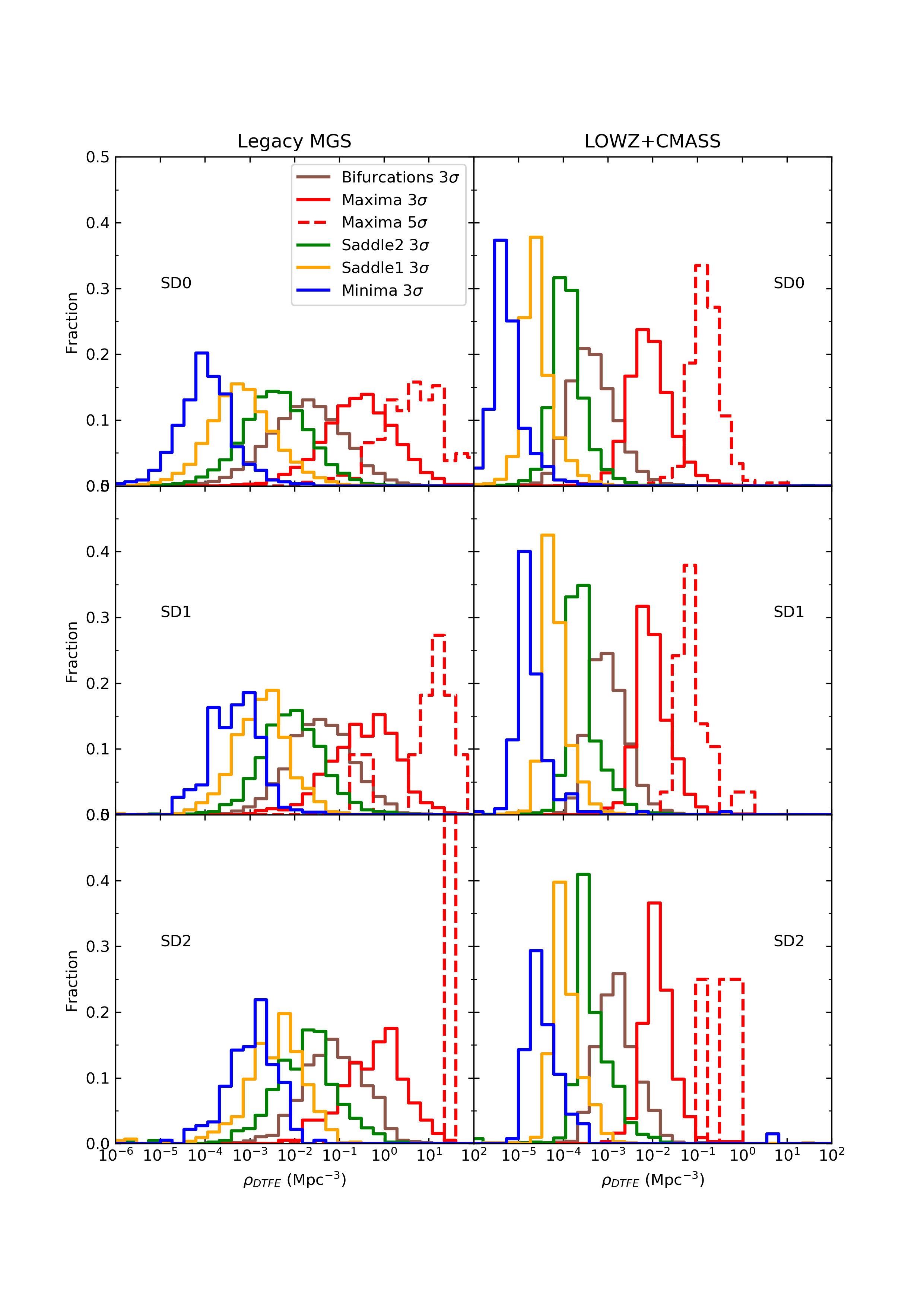}
\caption{Density distribution of the critical points as derived from the DTFE. The left and right columns refer to the Legacy MGS and LOWZ+CMASS sample, respectively. Rows are different smoothing cycles of the density field prior to filament detection. In each panel, the solid lines refer to maxima (red), bifurcations (brown), type 2 saddles (green), type 1 saddles (yellow), and minima (blue). Only the $3\sigma$ persistence threshold case is shown, except for maxima, where the $5\sigma$ persistence sample is also reported (dashed red lines). Although indicated as $5\sigma$, the high-persistence cut for LOWZ+CMASS in the one-smoothing case has been limited to $4.5\sigma$.}
\label{crit_dens}
\end{figure*}

Figure \ref{crit_dens} shows the DTFE density distribution of critical points divided by their type. This figure shows the wide diversity of the features of the density field identified by \disperse, as well as the broad range of densities explored. The density distributions of critical points overlap on a large area, which further confirms that a topological analysis is required to correctly identify cosmic structures (such as filaments) and not only a simple density criterion \citep[see also e.g.][]{Cautun2014}. The density distribution for bifurcation points is particularly interesting. These points are artificially inserted at the intersection position of two or more filaments and are intermediate between maxima of the density field and type 2 saddles, which are located in filaments. This particular feature makes them suitable candidates for being associated with low-mass groups or unresolved clusters.

Changing the level of smoothing reduces the range of densities explored, as expected, but does not significantly alter the shape of the distributions. Changing the persistence threshold, on the other hand, significantly affects the density distribution of maxima and eliminates the low-density tail of their distribution (dashed red line in the figures). The density distributions of the other types of critical points are left almost unaltered (not shown here for clarity).

We searched for a possible redshift evolution of these distributions and found that the absolute value of the DTFE density for the critical points decreases with increasing redshift. This behaviour is expected because the number of tracers is lower at higher redshift. However, the shape and partial overlap of the distributions do not change, and this is further confirmation that \disperse is able to correctly interpret the density field and detect the proper structures at all redshifts.

\subsection{Filaments}
\label{fil_cat_sec}
\begin{figure}
\centering
\includegraphics[width = \linewidth]{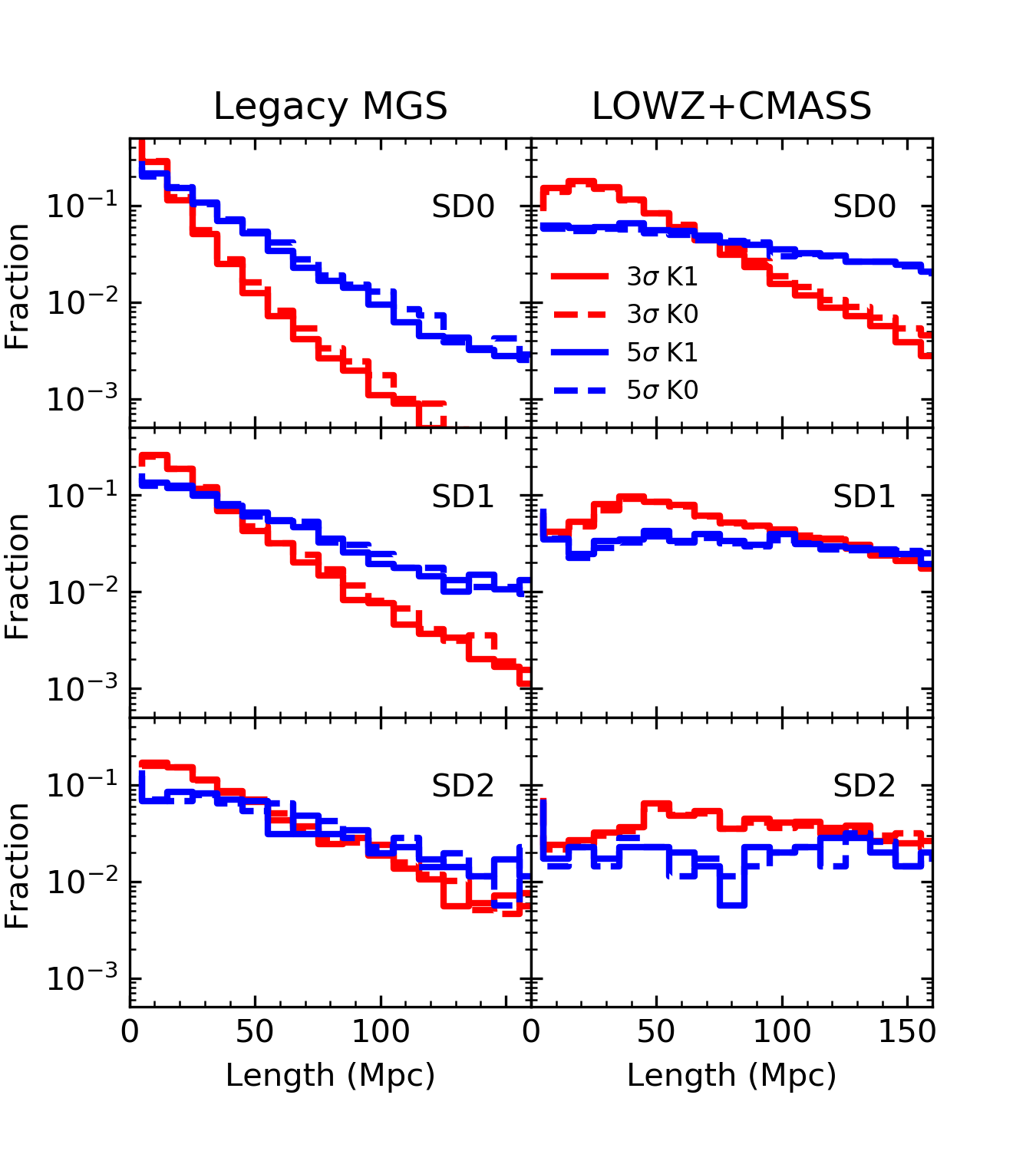}
\caption{Length distribution of the filaments. The left and right columns refer to the Legacy MGS and LOWZ+CMASS sample, respectively. The rows are different smoothing cycles of the density field prior to filament detection. The red lines refer to a $3\sigma$ persistence threshold, and blue lines refer to a $5\sigma$ persistence threshold. The dashed and solid lines represent filament samples when no smoothing of the skeleton is applied after filament identification (marked ``K0'' in the legend) and when one cycle of smoothing is applied (marked ``K1'' in the legend), respectively. Although indicated as $5\sigma$, the high-persistence cut for LOWZ+CMASS in the one-smoothing case has been limited to $4.5\sigma$.}
\label{length_dist}
\end{figure}

Figure \ref{length_dist} shows the length distribution of the filaments connecting maxima, bifurcations, and type 2 saddle points. Depending on the redshift distribution of the sample and on the density of available tracers, the distribution may be more peaked towards short (as in the Legacy MGS case) or towards longer filaments (e.g. LOWZ+CMASS). Figure \ref{length_dist} shows that increasing the persistence threshold essentially eliminates a large fraction of short, less significant filaments from the sample.

This figure also shows that a smoothing of the skeleton after its detection does not significantly alter the length distribution of the filaments. The skeleton-smoothing procedure allows removing unphysical edges and sharp turns in directions within the filaments that are likely due to the noise of the galaxy distribution. The effect is very weak, however and the filament properties are not significantly changed. 

\begin{figure*}
\centering
\includegraphics[width = \linewidth, trim = 0cm 2cm 1cm 2cm, clip = true]{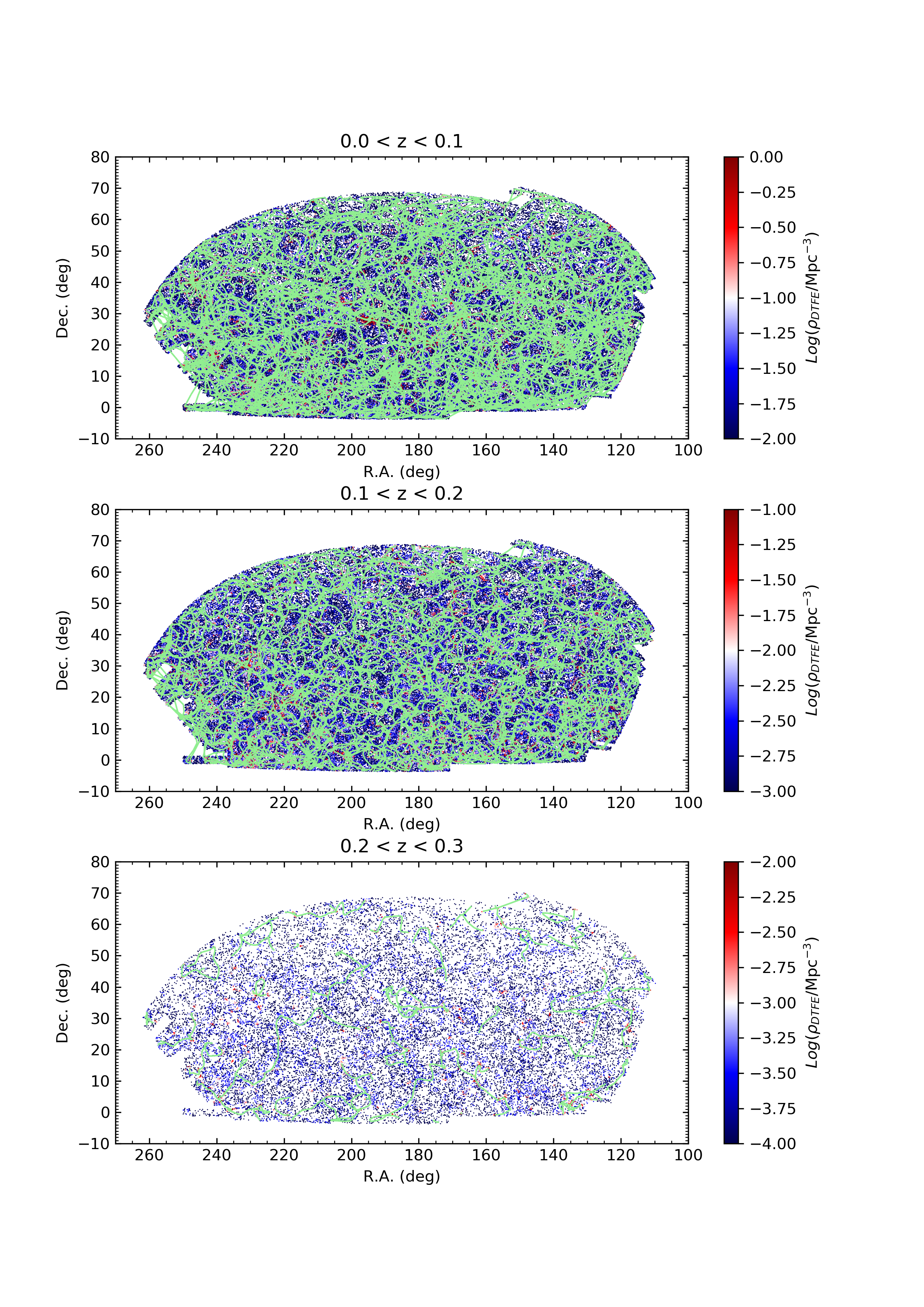}
\caption{Angular distribution of the filaments on the plane of the sky. The galaxy distribution is colour-coded according to the local density as measured by the DTFE, and filaments are over-plotted in green. Three ranges of redshift are considered for clarity, as marked at the top of each panel. Only the Legacy MGS is considered. As an example, filaments detected with a $3\sigma$ persistence threshold and one smoothing cycle of the density field have been reported. No cleaning of the skeleton is performed to show the full distribution of filaments as recovered by DisPerSE.}
\label{Fil_map_radec_legacy}
\end{figure*}

\begin{figure*}
\centering
\includegraphics[width = \linewidth, trim = 0cm 2cm 1cm 2cm, clip = true]{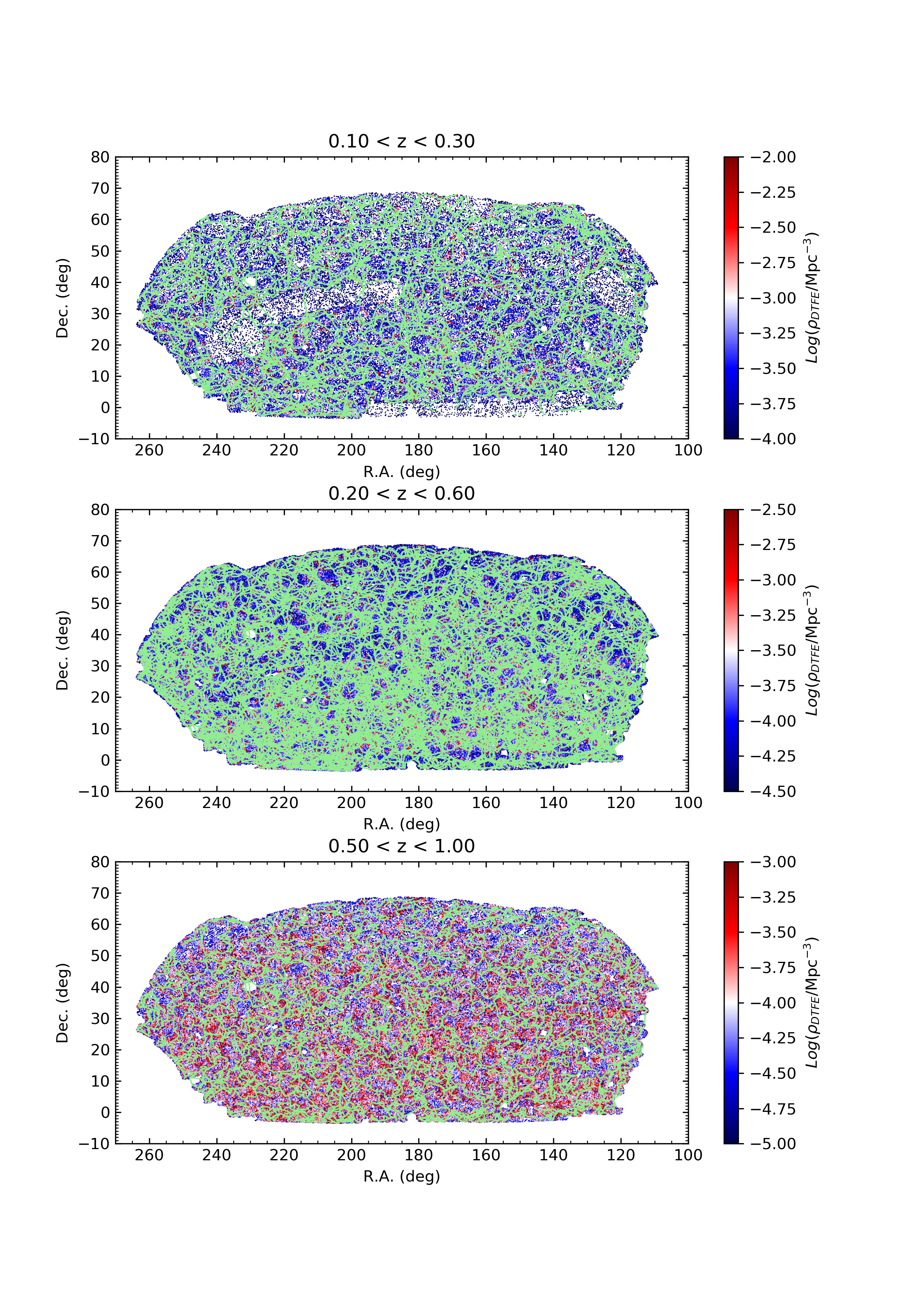}
\caption{Angular distribution of the filaments on the plane of the sky. The galaxy distribution is colour-coded according to their local density as measured by the DTFE, and filaments are over-plotted in green. Three ranges of redshift are considered for clarity, as marked at the top of each panel. Only the LOWZ+CMASS sample is considered. As an example, filaments detected with a $3\sigma$ persistence threshold and no smoothing of the density field have been reported. No cleaning of the skeleton is performed to show the full distribution of filaments as recovered by DisPerSE.}
\label{Fil_map_radec_lowzcmass}
\end{figure*}

Figures \ref{Fil_map_radec_legacy} and \ref{Fil_map_radec_lowzcmass} show the filament distribution in the plane of the sky (in three redshift bins for clarity, from top to bottom) for the Legacy MGS and LOWZ+CMASS skeletons, respectively. No cleaning of the skeleton for edge effects or other minor problems was performed to show the filaments as they are output by DisPerSE (figures \ref{Fil_map_radec_legacy_cleaned} and \ref{Fil_map_radec_lowzcmass_cleaned} in Appendix \ref{cfig} are versions of these figures that were made with the cleaned skeleton as an example). These figures give a visual representation of the filament distribution with respect to the galaxy distribution. Filaments clearly follow the galaxy density field, and dense nodes lie at their intersection and large low-density regions between them. These maps provide a full view of the LSS as detected in the volume of the Universe explored by the SDSS. Several defects are also visible in these maps: filaments are present where no galaxies are available, especially in the lowest redshift bin of the LOWZ+CMASS survey. This simultaneously highlights the power and limitations of the \disperse$\:$ method: the DTFE tetrahedrons are able to cross gaps, holes, and empty regions and essentially provide an interpolation of the density field across them. When the holes are too large (e.g. the lowest redshift bin of the LOWZ+CMASS survey, top panel of Figure \ref{Fil_map_radec_lowzcmass}), the interpolated density field is not reliable and the detected filaments are spurious.  \citet{Tanimura2019disperse} performed the analysis at $z > 0.2$ and discarded all the filaments at lower redshifts and outside of the footprint of the SDSS DR12.  \citet{Bonjean2019Filaments} also applied the SDSS DR12 mask, but they did not take the large holes in the footprint at low redshift ($z \lesssim 0.15$) into account. Still, given their size, in these regions the risk is rather not to detect any filaments than to detect spurious ones. This should decrease the significance of the signal for the filament profiles detected in the galaxy distribution rather than polluting it with spurious filaments because of a reduced statistics in terms of the number of available filaments. As the SDSS footprint has a large area and the number of filaments for the analysis is high, the conclusions derived in \citet{Bonjean2019Filaments} remain valid. Moreover, splitting the sample into two redshift bins ($0.1 < z < 0.2$ and $0.2 < z < 0.3$) yielded consistent results between the two, but with lower significance.

\begin{figure*}
\centering
\includegraphics[width = \linewidth]{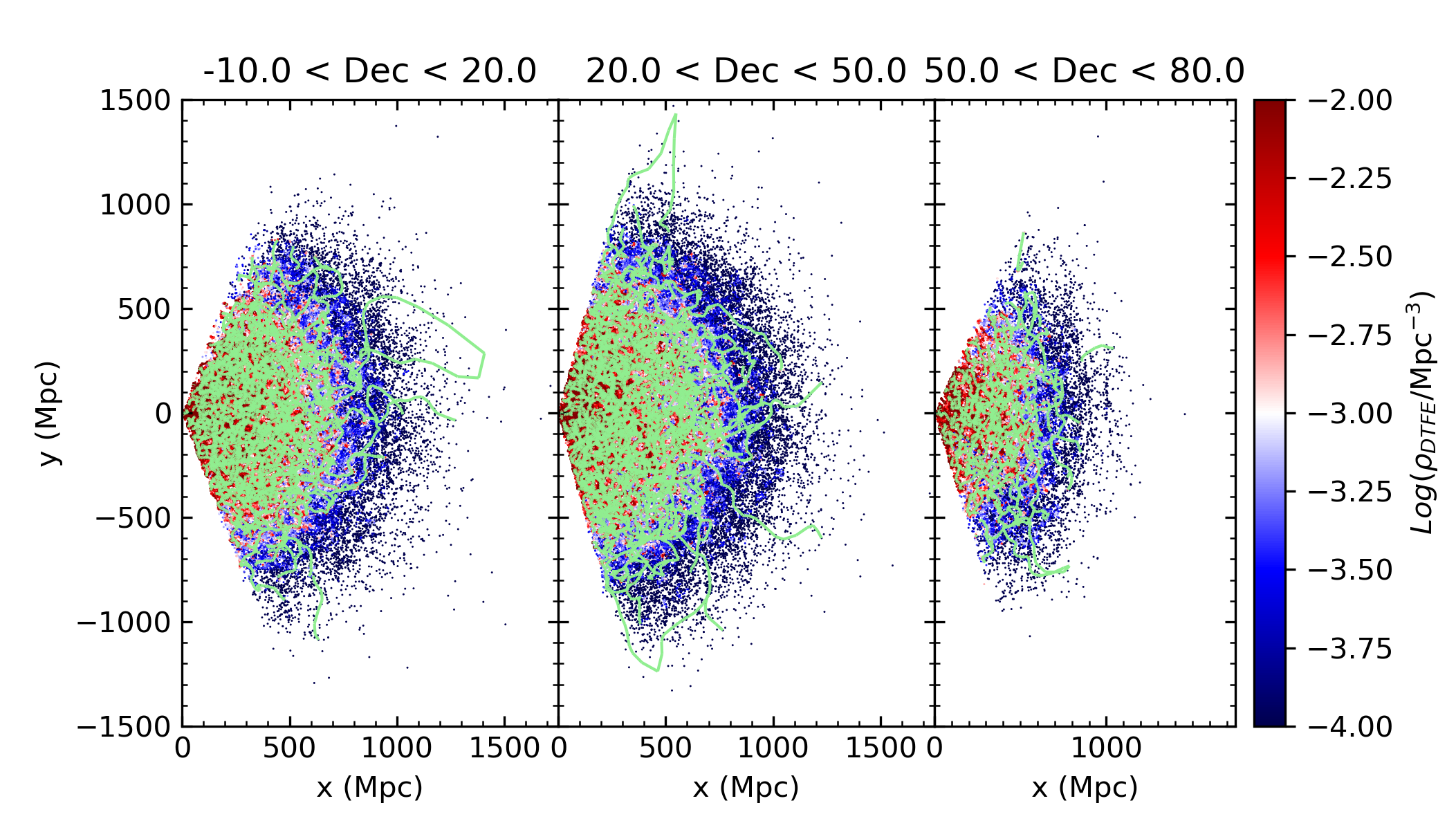}
\caption{Maps of filaments on a Cartesian $x-y$ plane. The $x$-axis is aligned with increasing redshift, and the $y$-axis with increasing right ascension. The galaxy distribution is colour-coded according to local density as measured by the DTFE, and filaments are over-plotted in green. Three ranges of declination are considered for clarity, as marked at the top of each panel. Only the Legacy MGS is considered. As an example, filaments detected with a $3\sigma$ persistence threshold and one smoothing cycle of the density field have been reported. No cleaning of the skeleton is performed to show the full distribution of filaments as recovered by DisPerSE.}
\label{Fil_map_raz_legacy}
\end{figure*}

\begin{figure*}
\centering
\includegraphics[width = \linewidth]{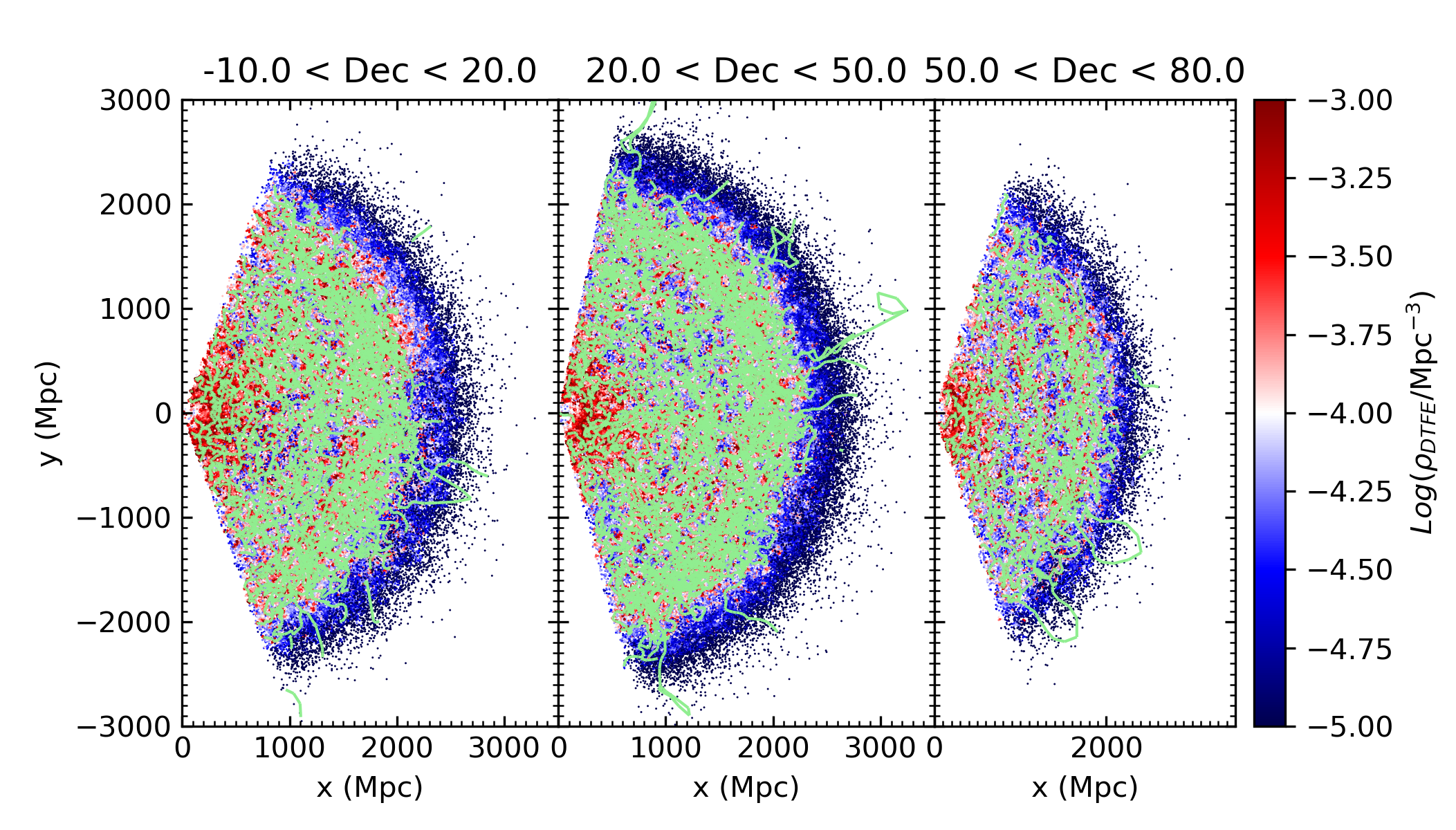}
\caption{Maps of filaments on a Cartesian $x-y$ plane. The $x$-axis is aligned with increasing redshift, and the $y$-axis with increasing right ascension. The galaxy distribution is colour-coded according to local density as measured by the DTFE, and filaments are over-plotted in green. Three ranges of declination are considered for clarity, as marked at the top of each panel. Only the LOWZ+CMASS is considered. As an example, filaments detected with a $3\sigma$ persistence threshold and no smoothing of the density field have been reported. No cleaning of the skeleton is performed to show the full distribution of filaments as recovered by DisPerSE.}
\label{Fil_map_raz_lowzcmass}
\end{figure*}

Figures \ref{Fil_map_raz_legacy} and \ref{Fil_map_raz_lowzcmass} show the filament distribution in a transverse cut in a Cartesian reference system, aligned with the redshift direction ($x$-axis) and with right ascension ($y$-axis), in three declination bins for clarity (from left to right). No cleaning of the skeleton for edge effects or other minor problems was performed here either (figures \ref{Fil_map_raz_legacy_cleaned} and \ref{Fil_map_raz_lowzcmass_cleaned} in Appendix \ref{cfig} are versions of these figures made with the cleaned skeleton as an example). These maps show that the filaments we detect fill the entire survey volume, following the galaxy density field, and do not have a preferential angular direction or a strong trend with redshift. These maps also show several defects: long filaments extend from the edge of the survey to regions that are almost devoid of galaxies; these filmanets are several hundred megparsec long. They are probably due to spurious critical points created at the edge of the survey volume. The cuts in length of the filaments that were applied to the samples in \citet{Tanimura2019disperse} and \citet{Bonjean2019Filaments} serve the purpose of eliminating these features.  

\begin{figure}
\centering
\includegraphics[width = \linewidth]{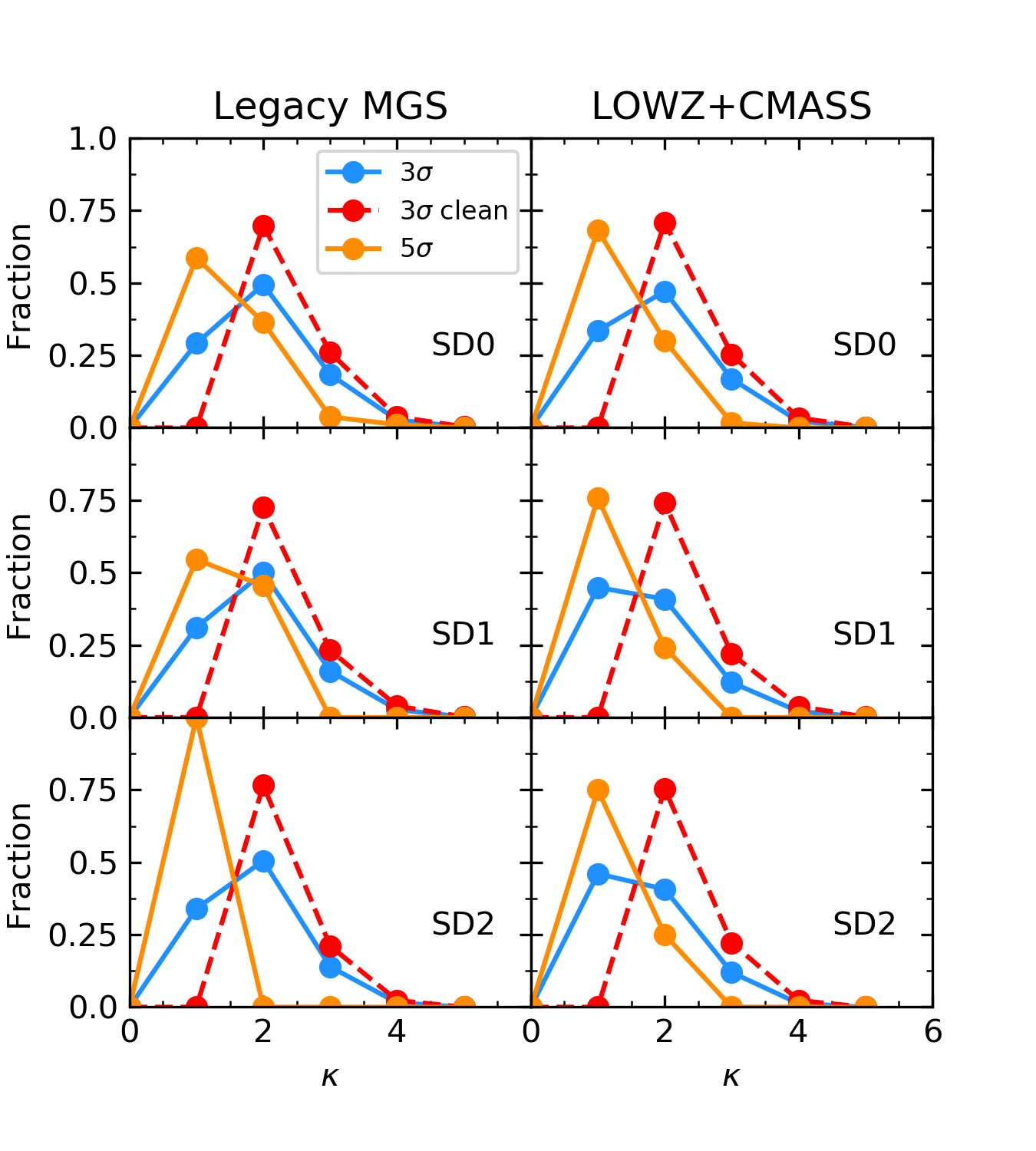}
\caption{Connectivity distributions for maxima. The left and right columns refer to the Legacy MGS and LOWZ+CMASS sample, respectively. The rows are different smoothing cycles of the density field prior to filament detection. The cyan lines refer to a $3\sigma$ persistence threshold, and orange lines refer to a $5\sigma$ persistence threshold. The dashed red lines are connectivity distributions for a $3\sigma$ persistence threshold after maxima were removed that were connected to only one filament. Although indicated as $5\sigma$, the high-persistence cut for LOWZ+CMASS in the one-smoothing case has been limited to $4.5\sigma$.}
\label{crit_connectivity}
\end{figure}

Figure \ref{crit_connectivity} shows the distributions of the number of filaments that is connected to maxima. This quantity is called connectivity, and it is an important observable for the study of the cosmic web, of structure formation, and cosmology \citep[see e.g.][]{Codis2018, Sarron2019, DarraghFord2019, Malavasi2019, Kraljic2019connectivity}. The connectivity of maxima scales with the density of the maxima, that is, with the mass of the clusters associated with the peaks of the density field, and it is usually a number in the range $0 \div 10$. Figure \ref{crit_connectivity} shows that the connectivity values are consistently distributed in the range $0 \div 3$  for the different smoothing levels and the different samples. Increasing the persistence threshold has the effect of decreasing the average connectivity, as more filaments are eliminated from the sample.

We note that the notion of a maximum connected to only one filament may be considered unphysical because not many isolated maxima are expected in a fully connected skeleton. In all the panels of Figure \ref{crit_connectivity} we report the distributions of connectivity values for the maxima when we removed the 0 Mpc filaments and the maxima with only one filament for a $3\sigma$ persistence. As expected, the connectivity distributions now peak at a value of two filaments connected to maxima. Of particular interest is the fact that in the $5\sigma$ persistence threshold and two smoothing cycles for the Legacy MGS all maxima are connected to one filament, therefore no maxima are left in the sample after the cleaning. This contributes to better understanding the best combination of \disperse parameters to use to detect the filaments in this particular galaxy sample and further discourages the use of such a high persistence threshold combined with a high degree of smoothing in this particular case.

\section{Validation of the catalogue}
\label{validation}

In this section we analyse a few of the possible systematic effects that can affect the reconstruction of the cosmic web, as well as possible ways to handle them. In particular, we focus on two main items: the effect of the so-called Finger-of-God (FoG) redshift distortions \citep{Jackson1972,SargentTurner1977,Kaiser1987}, and the edge effects that are due to the border of the survey.

\subsection{Finger-of-God redshift space distortions}
The FoG effect is a distortion in the position of galaxies along the LoS that is caused by the peculiar motions inside clusters. When the redshift of a source is measured, the observed redshift is the sum of the intrinsic cosmological redshift and of a component due to the peculiar velocities of galaxies inside structures. As redshift is a measurement of the velocity component along the LoS, structures are not distorted in the directions that lie on the plane of the sky. As a result, clusters appear elongated along the LoS, and their symmetry is more cylindrical than spherical. This may present a problem, as an algorithm like \disperse may misinterpret clusters distorted by the FoG effect for filaments. This would affect the subsequent analysis because very many straight filaments would be aligned along the LoS.

\begin{figure*}
\centering
\includegraphics[width = \linewidth, trim = 1cm 1cm 1cm 1cm, clip = true]{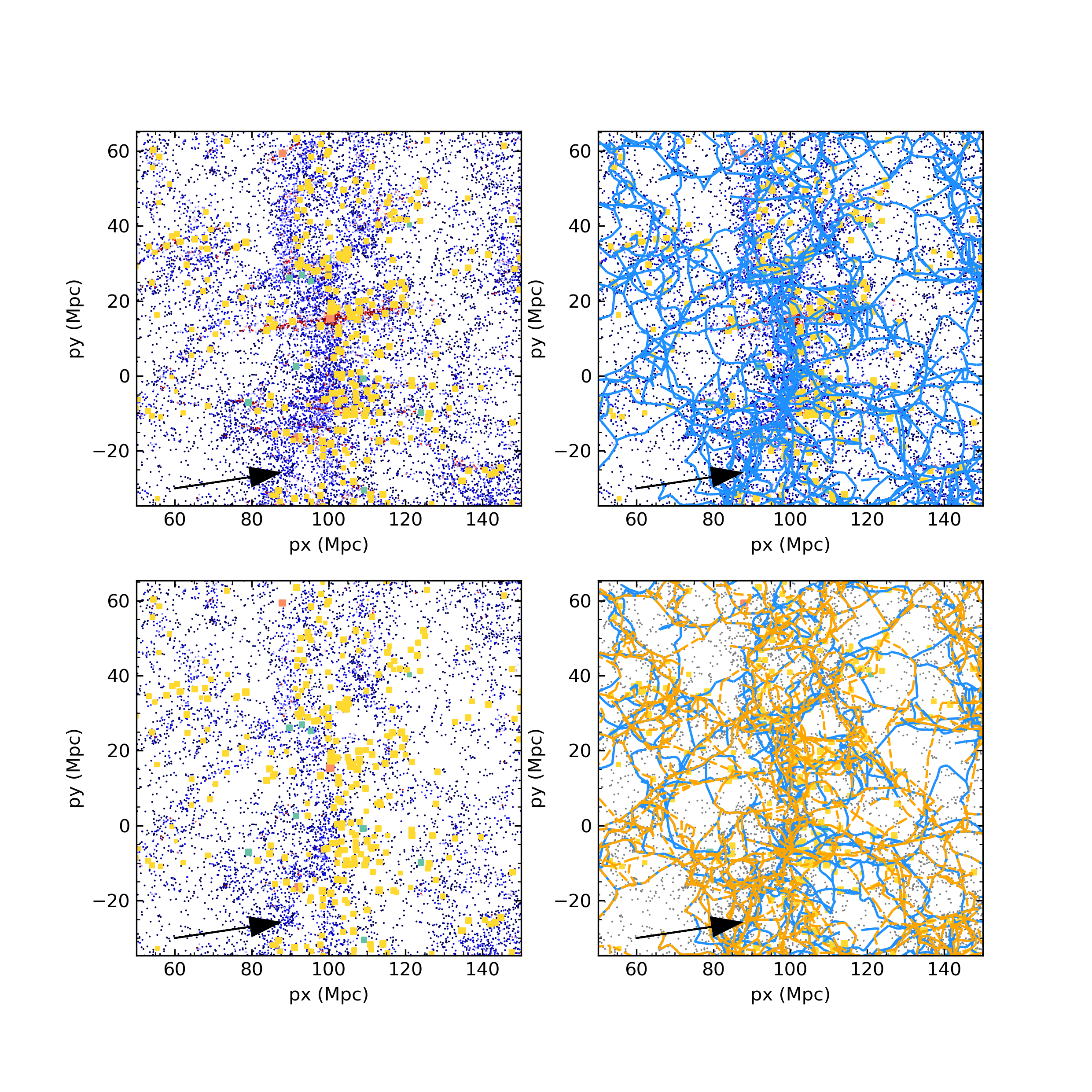}
\caption{Effect of FoG on the skeleton reconstruction around Coma. The four panels show a $50 \times 50 \times 50$ Mpc$^{3}$ box centred on the Coma cluster, projected on the Cartesian $x-y$ plane (the $x$-axis is aligned with redshift and the $y$-axis with right ascension). In all panels small points correspond to galaxies from the Legacy MGS. In three out of four panels they are colour-coded according to local density. In the top panels, the FoG effect is present in the galaxy distribution, but it has been removed in the bottom panels. In the right panels, the blue skeleton has been derived with the FoG effect, and the dashed orange skeleton has been derived after the FoG effect was removed. Both skeletons have been derived with no smoothing of the density field and a $3\sigma$ persistence threshold. In each panel, squares correspond to the clusters we used to eliminate the FoG effect, colour-coded according to the sample to which they belong (not all samples have clusters inside the box; MCXC X-ray clusters are plotted in aqua, Planck SZ clusters in orange, and \citealt{Tempel2017} optical clusters in yellow). The size of the squares scales with cluster virial mass. In each panel the black arrow shows the direction of the LoS.}
\label{fog_coma}
\end{figure*}

We addressed this problem by modifying the tessellation and re-running \disperse. This eliminated galaxies that might be affected by the FoG problem. Figure \ref{fog_coma} shows the steps adopted by our procedure, which we summarise below. This figure shows a cubic region of 50 Mpc a side, centred on the Coma cluster. This region was taken as an example both because it was the subject of the analysis of \citet{Malavasi2019} and because it is a massive cluster with a very prominent distortion. The top left panel shows the galaxies from the Legacy MGS colour-coded according to their density as computed from the DTFE. The FoG distortion due to Coma is clearly visible as an elongated structure (in all four panels, a black arrow indicates the direction of the LoS). In the same figure, coloured squares mark the position of clusters coming from different samples (see Section \ref{subsectionclusters} of this work and Section 2 of \citealt{Malavasi2019}).

To address the problem of the FoG effect in the skeleton reconstruction, we adopted the following steps:
\begin{enumerate}
\item We centred a cylinder on each cluster and identified all the galaxies inside. The cylinder had a radius $R = 3 \times R_{\mathrm{vir}}$ and height $\mathrm{d}z = \pm 3 \times (\sigma_{v}/c) \times (1+z),$ where $\sigma_{v}$ is the velocity dispersion computed using the virial mass as $\sigma_{v} = \sqrt{GM_{\mathrm{vir}}/5R_{\mathrm{vir}}}$.
\item The galaxies inside the cylinders centred on each cluster from the samples mentioned above were removed from the galaxy catalogue. At the same time, a new entry was added to the galaxy catalogue at the position of each cluster. To each of these new particles, a density was given equal to the average of the DTFE densities of the galaxies inside the cylinder. The densities of all other galaxies were left unaltered.
\item The tessellation was recomputed to properly rebuild the tetrahedrons on which the density field function is defined for \disperse to run properly. The DTFE density was not recomputed, but previous values were used for galaxies that were not removed, and the new average values computed inside the cylinders were used for the new particles that were added at the position of clusters. This avoids the formation of spurious minima at the positions of clusters where holes are left by the removal of galaxies in the cylinders.
\item \disperse was run as normal on the new tessellation with the modified density field.
\end{enumerate}
This method essentially changes the shape of the density field locally but leaves the density values rather unaltered. The bottom left panel of Figure \ref{fog_coma} shows the galaxy distribution after the removal of the FoG cylinder, with the same colour-coding as the top left panel. The elongated feature at the Coma position is clearly missing. The top right panel shows the skeleton as detected in the galaxy distribution with the FoG at the position of Coma, and the bottom right panel shows a comparison between the skeleton at the position of Coma computed with the FoG effect and when the FoG effect was removed from the galaxy distribution using the above procedure.

When we compute the skeleton with the FoG distortion, a filament is clearly visible in the LoS direction, at the position of the Coma cluster. When the FoG distortions are removed, the filament along the LoS detected at the position of Coma is still present, although its shape has changed. This confirms that although the distortion of the FoG effect affects the skeleton, the detection of filaments at the position of clusters, even massive ones such as Coma, can still be considered reliable. The bottom right panel of Fig. \ref{fog_coma} shows that the skeleton has also changed in several other positions because many other clusters also lie in the region, but the skeleton that we re-computed after the distortions from the FoG effect were removed is still visually similar to the original.

\begin{figure*}
\centering
\includegraphics[width = \linewidth, trim = 1cm 1cm 1cm 0cm, clip = true]{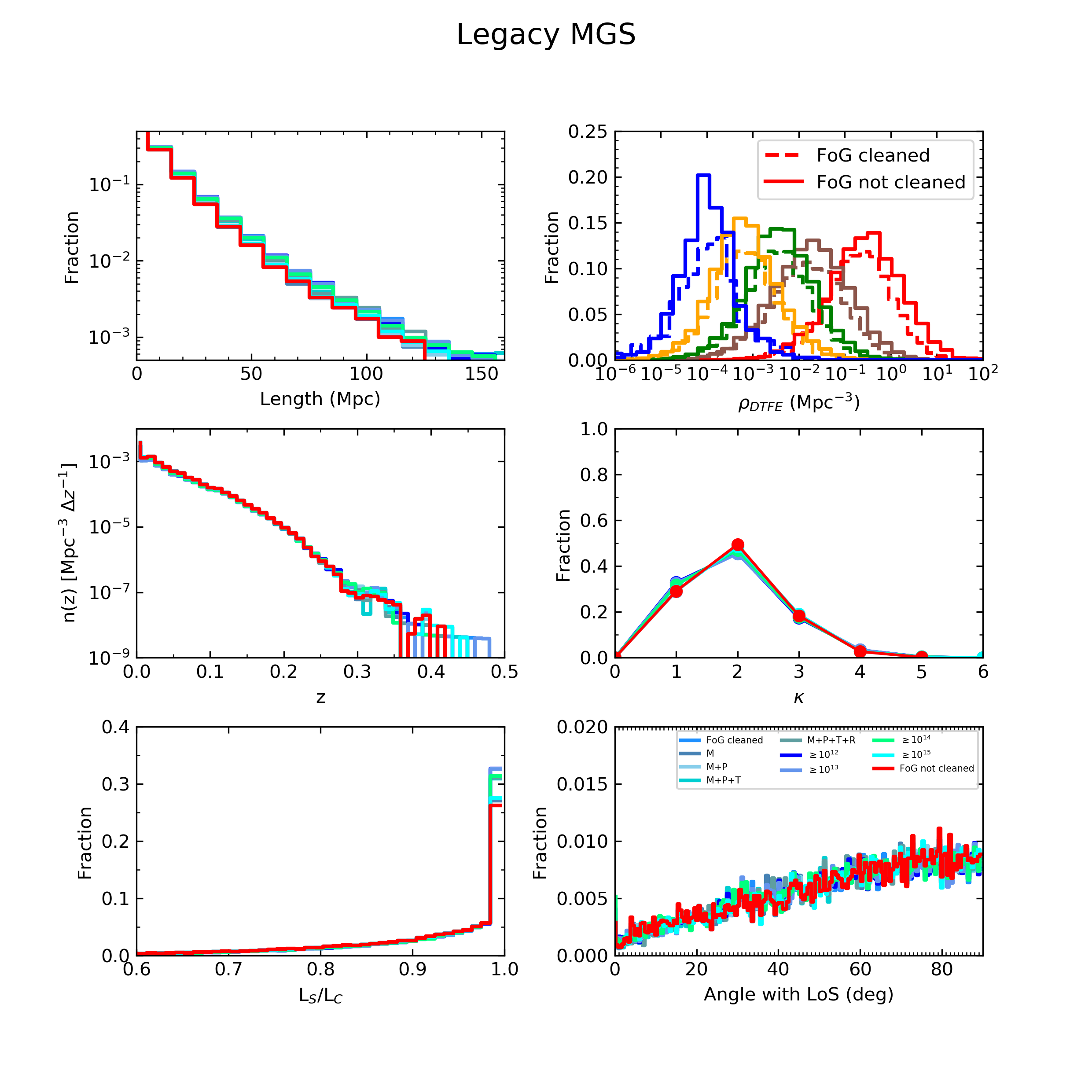}
\caption{Statistical effects of the FoG effect on the skeleton. The six panels show the distributions of the filament lengths (top left), of the density of critical points divided by type (top right), of the number density of the critical points as a function of redshift (middle left), of the connectivity of the maxima (middle right), of the curvature of the filaments (bottom left), and of the angle with the LoS of the straight filaments (bottom right). In the top right panel, the distributions are colour-coded according to the critical point type they refer to (blue: minima, yellow: type 1 saddles, green: type 2 saddles, red: maxima, brown: bifurcations). The solid lines refer to the distributions obtained for the case where the skeleton has been derived without the FoG effect and the dashed lines refer to the case where the FoG effect is included. In all the other panels, the red line refers to the case where the FoG is included in the derivation of the skeleton, and the different shades of cyan show the case where the FoG effect is removed using different samples of clusters or clusters of different mass, as explained in the text. In the legend, ``M'' refers to MCXC X-ray clusters, ``P'' to Planck SZ clusters, ``T'' and ``R'' to \citet{Tempel2017} and RedMaPPer optical clusters, respectively, and ``FoG cleaned'' to the case where all the clusters are used. This figure is for the Legacy MGS case.}
\label{fog_histograms}
\end{figure*}

\begin{figure*}
\centering
\includegraphics[width = \linewidth, trim = 1cm 1cm 1cm 0cm, clip = true]{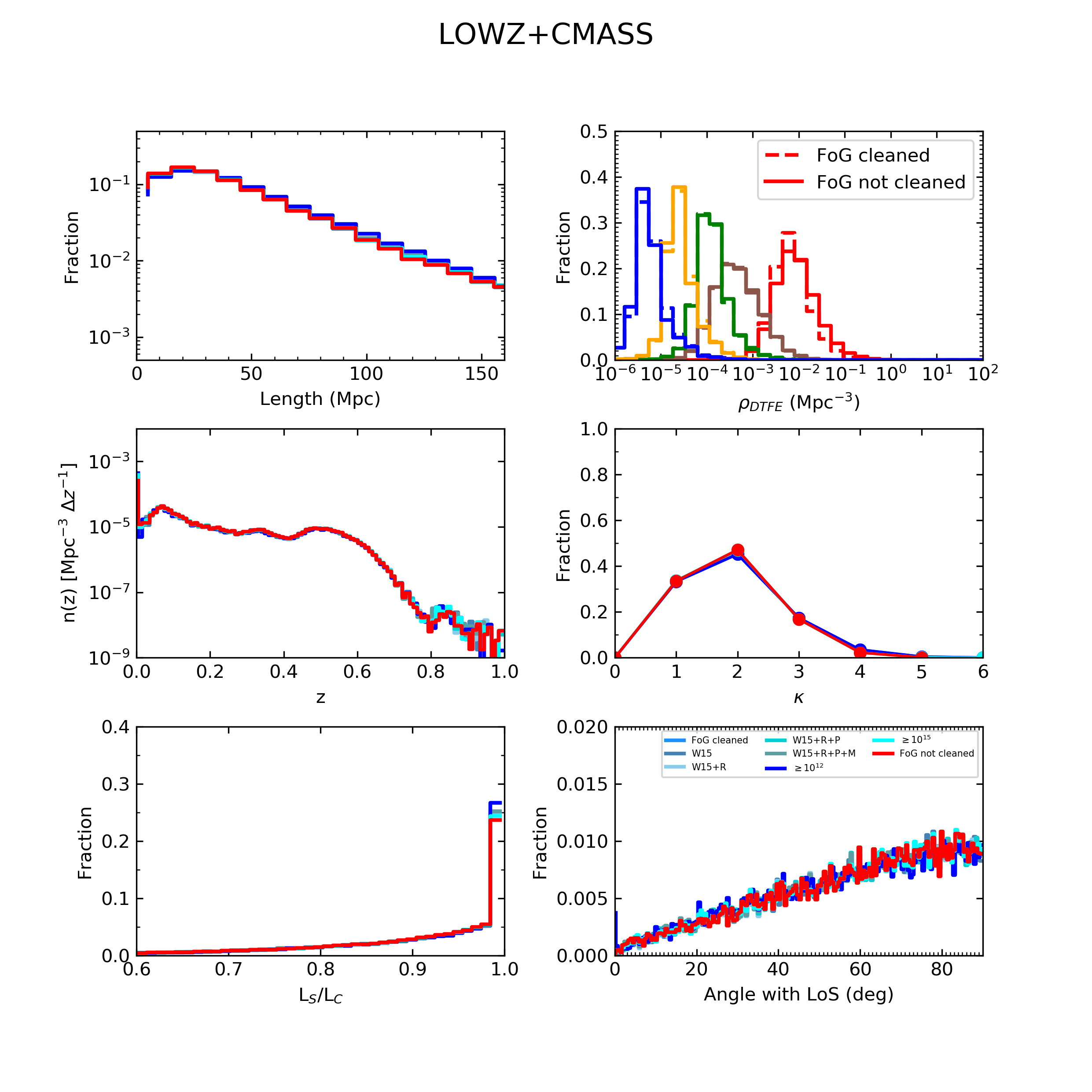}
\caption{Statistical effects of the FoG effect on the skeleton. The six panels show the distributions of the filament lengths (top left), of the density of critical points divided by type (top right), of the number density of the critical points as a function of redshift of the critical points (middle left), of the connectivity of the maxima (middle right), of the curvature of the filaments (bottom left), and of the angle with the LoS of the straight filaments (bottom right). In the top right panel, distributions are colour-coded according to the critical point type they refer to (blue: minima, yellow: type 1 saddles, green: type 2 saddles, red: maxima, brown: bifurcations). The dashed lines refer to the distributions obtained for the case where the skeleton has been derived without the FoG effect and solid lines refer to the case where the FoG effect is included. In all the other panels, the red line refers to the case where the FoG is included in the derivation of the skeleton, and the different shades of cyan show the case where the FoG effect is removed using different samples of clusters or clusters of different mass, as explained in the text. In the legend, ``M'' refers to MCXC X-ray clusters, ``P'' to Planck SZ clusters, ``W15'' and ``R'' to \citet{Wen2015} and RedMaPPer optical clusters, respectively, and ``FoG cleaned'' to the case where all the clusters are used. This figure is for the LOWZ+CMASS case.}
\label{fog_histograms_lowzcmass}
\end{figure*}

We investigated how the removal of the FoG distortions affects the statistical properties of the skeleton. We produced the distributions of filament lengths, the number density of critical points as a function of redshift, the density of critical points divided by type, and of the connectivity of the maxima in the case with and without distortions from the FoG effect (visible in the upper four panels of Figs. \ref{fog_histograms} and \ref{fog_histograms_lowzcmass} for the Legacy MGS and LOWZ+CMASS samples, respectively). These distributions are very similar in the two cases, which shows that statistically, there is no difference between the two sets of filaments. To further test the effect of distortions caused by the FoG effect, we computed the level of curvature of the filaments. This quantity is defined as the ratio between the length of the filaments measured as the length of the straight line connecting the critical points at the filament extremities ($L_{S}$) over the length of the filaments computed by summing the length of the individual segments that compose each filament and follow the filament path ($L_{C}$). Filaments with a ratio ($L_{S}/L_{C}$) close to one will be straight, meaning that the filament path does not differ too much from a straight line that connects the filament extremities. If $L_{S}/L_{C} \ll 1$, it means that $L_{C} \gg L_{S}$ , and the filament will be curved. While straight filaments are perfectly possible even in the perfect case of a galaxy distribution without distortions caused by the FoG effect, their incidence should be higher in a distorted density field. Figures \ref{fog_histograms} and \ref{fog_histograms_lowzcmass} show that the distributions of filament curvature are very similar regardless of whether we remove distortions caused by the FoG effect. 

We further analysed the statistical properties of the skeleton by choosing only the filaments with $L_{S}/L_{C} > 0.99$, that is, those for which the length measured following the filament is less than $1\%$ longer than a simple straight line that connects the extremities. For these very straight filaments, we computed the angle with respect to the LoS, defined as the angle between the straight line connecting the observer with the closer of the two critical points defining the filament and the straight line that connects the two filament extremities. This angle has values in the range $0 \div 90 \deg,$ and we show the distribution of the angles with LoS for straight filaments in the skeletons detected with and without the FoG distortions in the bottom right panels of Figs. \ref{fog_histograms} and \ref{fog_histograms_lowzcmass}. Small differences are visible between the cases with or without FoG distortions, but filaments are not preferentially aligned perpendicular to the LoS in either case. On the other hand, the distribution is rather uniform, showing that filaments have random orientations with respect to the LoS.

As a final search for potential biases that are due to the samples of clusters we used, we derived the distributions of Figures \ref{fog_histograms} and \ref{fog_histograms_lowzcmass} several times, using an increasing number of clusters each time. In particular, for the Legacy MGS we derived the distributions by first considering only the MCXC X-ray cluster sample and then adding in sequence the {\it Planck} SZ clusters, \citet{Tempel2017} and RedMaPPer optical clusters, and finally, the \citet{Wen2012} sample. For the LOWZ+CMASS sample, we started with only the \citet{Wen2015} clusters, followed by the RedMaPPer clusters, the Planck SZ clusters, the MCXC X-ray clusters, and finally, the \citet{Wen2012} sample. The different order for the cluster samples considered in the Legacy MGS and LOWZ+CMASS case is due to the different redshift ranges that are covered by the galaxy surveys and the cluster samples. We also performed the FoG analysis by considering all cluster samples, but with different mass thresholds: for the Legacy MGS, all clusters with $M_{vir} \geq 10^{12}$, $M_{vir} \geq 10^{13}$, $M_{vir} \geq 10^{14}$, and $M_{vir} \geq 10^{15}$. For the LOWZ+CMASS sample, the two intermediate-mass bins ($M_{vir} \geq 10^{13}$ and $M_{vir} \geq 10^{14}$) resulted in a crash of \disperse, probably because the removal of the FoG regions modifies the shape of the density field in such a way that it prevents \disperse from computing its gradient. We therefore considered only the two extreme ones (i.e. $M_{vir} \geq 10^{12}$ and $M_{vir} \geq 10^{15}$), which still allowed us to detect any trend in the FoG effect on the skeleton with cluster mass. We report the resulting distributions in Figures \ref{fog_histograms} and \ref{fog_histograms_lowzcmass}. No trend with the cluster sample adopted or with the mass threshold are visible, except for a slight increase in the incidence of straight filaments for some of the combinations of cluster samples.

When we performed this analysis, we were aware that some clusters can be present in more than one catalogue, either at the same position and with the same properties (virial radius, mass, and velocity dispersion) or at slightly different positions and with different properties. We did not specifically search for duplicate clusters in the various catalogues, as we decided to take the conservative approach of eliminating galaxies in the vicinity of each cluster from the sample. In case of a mismatch in the position of a cluster in two or more catalogues, a cylinder was placed at the cluster position each time and galaxies in a larger region were eliminated. Moreover, when for the same cluster different radii and masses are given in the catalogues, the highest values were taken consistently throughout the analysis, and galaxies were again eliminated from a larger region around the cluster position.

\subsection{Edge effects}
\begin{figure}
\centering
\includegraphics[width = \linewidth]{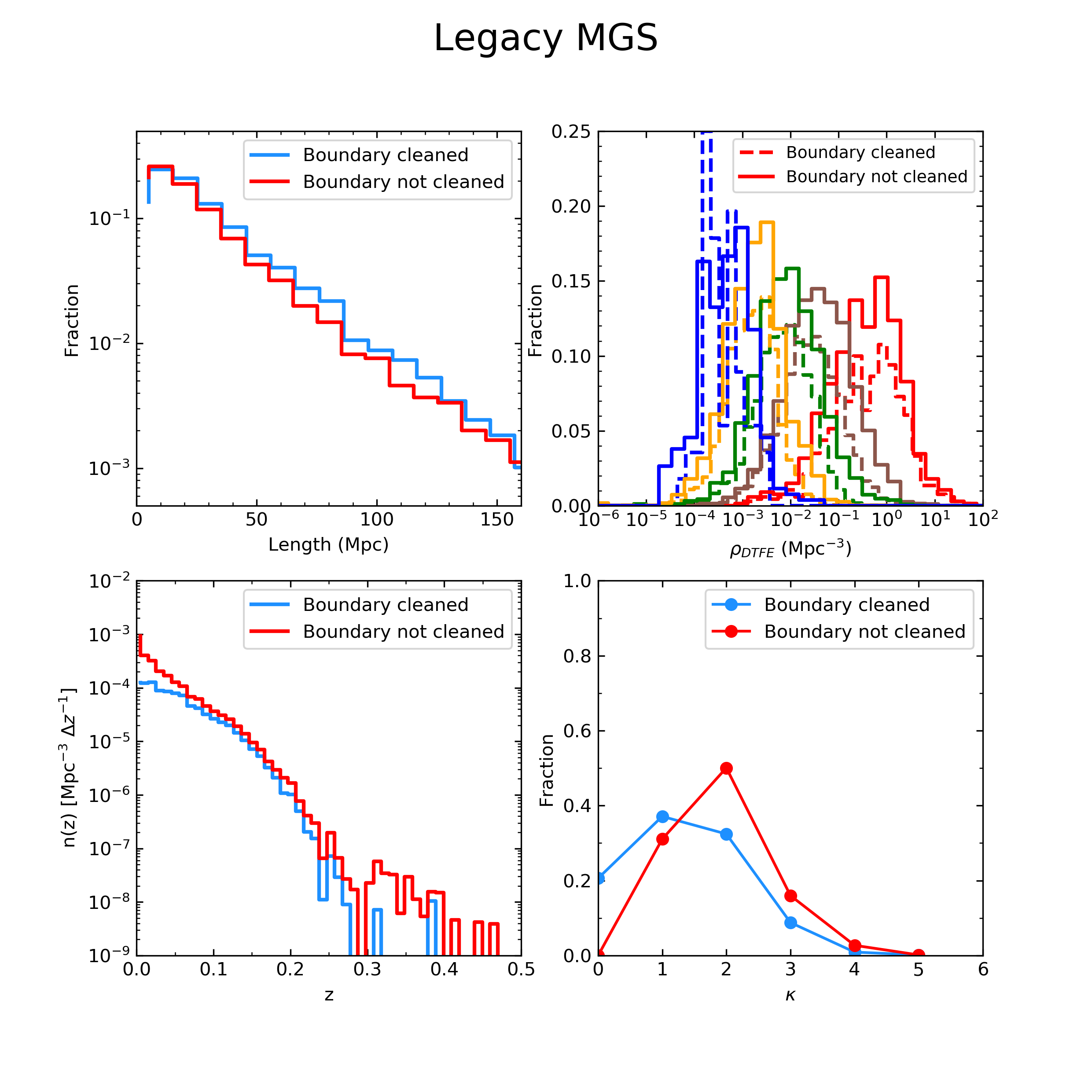}
\caption{Distributions of skeleton quantities when boundary effects are taken into account. The four panels show the distributions of the filament lengths (top left), of the density of critical points divided by type (top right), of the number density of critical points as a function of redshift (bottom left), and of the connectivity of the maxima (bottom right). In the top right panel, distributions are colour-coded according to the critical point type they refer to (blue: minima, yellow: type 1 saddles, green: type 2 saddles, red: maxima, brown: bifurcations). The dashed lines refer to the distributions obtained for the case where the critical points at the boundary and the filaments connected to them have been removed, and solid lines show when they have been included. In all the other panels, the cyan line refers to the case where the critical points at the boundary and the filaments connected to them have been removed, and the red line shows the case where they have been included. This figure is for the Legacy MGS case.}
\label{boundary_histograms}
\end{figure}

\begin{figure}
\centering
\includegraphics[width = \linewidth]{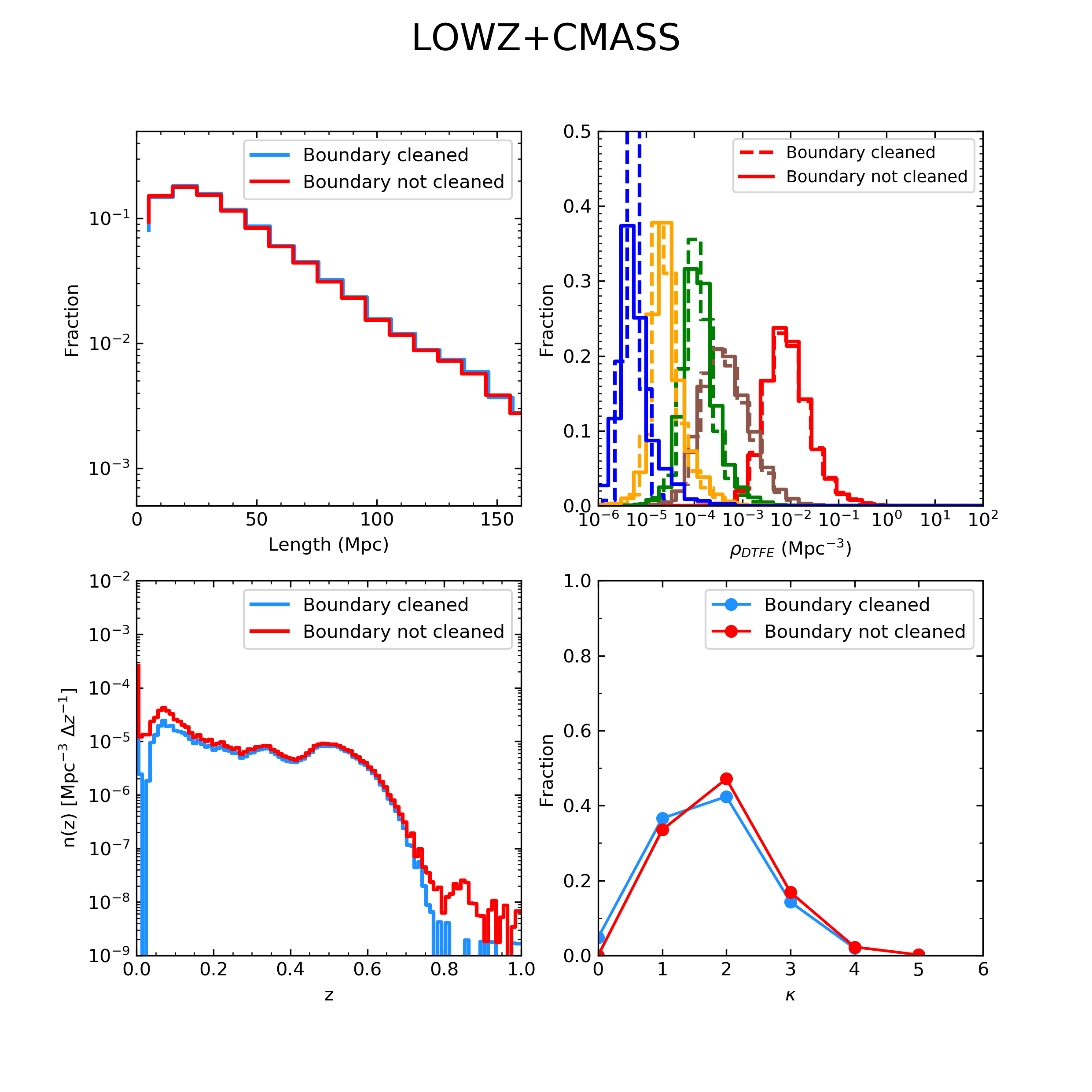}
\caption{Distributions of skeleton quantities when boundary effects are taken into account. The four panels show the distributions of the filament lengths (top left), of the density of critical points divided by type (top right), of the number density of critical points as a function of redshift (bottom left), and of the connectivity of the maxima (bottom right). In the top right panel, distributions are colour-coded according to the critical point type they refer to (blue: minima, yellow: type 1 saddles, green: type 2 saddles, red: maxima, brown: bifurcations). The dashed lines refer to the distributions obtained for the case where the critical points at the boundary and the filaments connected to them have been removed, and solid lines show when they have been included. In all the other panels, the cyan line refers to the case where the critical points at the boundary and the filaments connected to them have been removed, and the red line shows the case where they have been included. This figure is for the LOWZ+CMASS case.}
\label{boundary_histograms_lowzcmass}
\end{figure}

Figures \ref{boundary_histograms} and \ref{boundary_histograms_lowzcmass} show the same distributions of filament lengths, critical point number density as a function of redshift, critical point density for different types, and connectivity of maxima for the case of a skeleton where the critical points on the boundary have been kept and one where they have been removed, as described in Section \ref{method_edgeeffects}. Removing the critical points and filaments on the boundary has the effect of removing short filaments and of shifting the connectivity distributions for maxima to lower values. Nevertheless, the change in the statistical properties of the skeleton is not dramatic overall, especially for the redshift distribution of the number density of critical points and for the density distributions of critical points divided by type.

\subsection{Large-scale variations in the galaxy distribution}
\begin{figure}
\centering
\includegraphics[width = \linewidth]{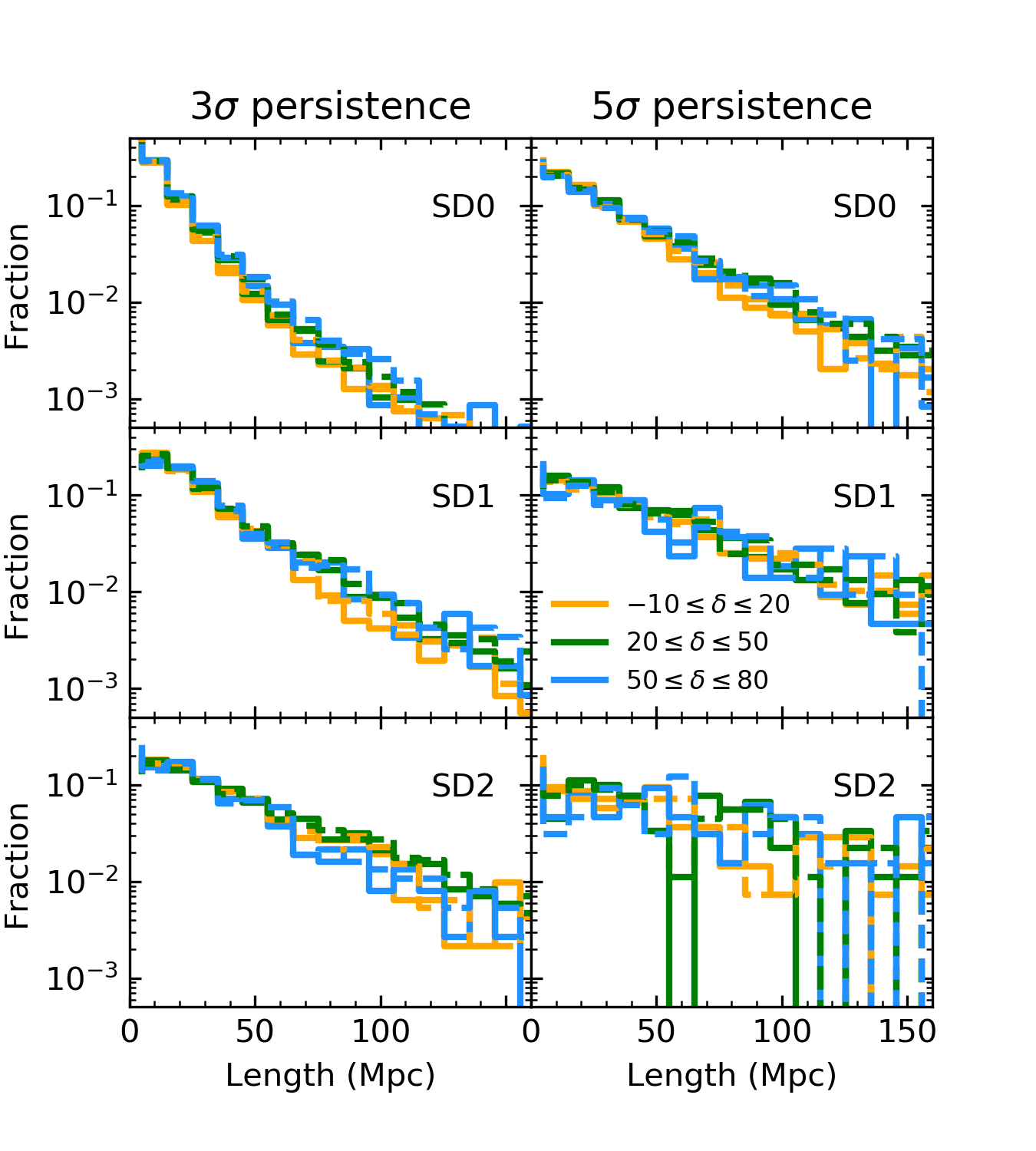}
\caption{Length distribution of the filaments in three declination slices. The left and right columns refer to the two persistence thresholds we used to derive the skeleton for the Legacy MGS, i.e. $3\sigma$ and $5\sigma$. The rows show different smoothing cycles of the density field prior to filament detection. The three declination slices are $-10 \leq \delta \leq 20 \deg$ (yellow), $20 \leq \delta \leq 50 \deg$ (green), and $50 \leq \delta \leq 80 \deg$ (cyan). Dashed lines and solid lines represent filament samples when no smoothing of the skeleton is applied after filament identification and when one cycle of smoothing is applied, respectively.}
\label{length_dist_slice}
\end{figure}

As a final test, we searched for the effect of potential large-scale inhomogeneities in the galaxy distribution on the plane of the sky on the reconstruction of filaments in the Legacy MGS. In particular, the surface density of galaxies appears to show a small but detectable trend with declination in Figure \ref{angdist_legacy}. To search for possible systematics, we derived the length distributions in three declination slices, which we report in Figure \ref{length_dist_slice}. The length distributions of the filaments in the three declination slices look identical regardless of the adopted persistence and smoothing level. Combined with visual inspection of the filaments shown in Figure \ref{Fil_map_raz_legacy}, this allows us to exclude any systematic trend in the reconstruction of the skeleton with declination.

\subsection{Qualitative comparison with the literature}
In recent years, other catalogues of filaments in the SDSS have been created using various methods to detect the cosmic web. For example, \citet{Tempel2014} used the BISOUS algorithm on the full SDSS Legacy MGS footprint, \citet{Martinez2016} detected filaments in the SDSS Legacy MGS by considering pairs of groups of galaxies, while \citet{Chen2016SDSS} extracted the cosmic web in 2D redshift slices using the SCMS algorithm on both the Legacy MGS and the LOWZ+CMASS sample. While a comparison of these catalogues is beyond the goal of this work \citep[see e.g.][]{Rost2019} also because very different algorithms were employed, we still qualitatively compare our work with others from the literature. While the sky distribution of filaments appear to be visually similar (see e.g. Figure 2 of \citealt{Chen2016SDSS}, Figure 8 of \citealt{Tempel2014} and Figure \ref{Fil_map_radec_legacy} of this paper), the length distributions reported for example in \citet[][Figure 4]{Rost2019} and in \citet[][Figure 11]{Tempel2014} show that these other catalogues find much shorter filaments than those we detect. Length distributions for the \citet{Tempel2014} and \citet{Martinez2016} catalogues encompass the range $0 \div 40$ Mpc, with the \citet{Martinez2016} catalogue having shorter filaments, in the range $0 \div 15$ Mpc. Figure \ref{length_dist} of this work shows that our filaments can reach lengths up to 100 Mpc, and the distributions vary according to the smoothing of the density field and persistence threshold. Moreover, our skeleton provides topologically motivated information on the critical points (maxima, bifurcations, and saddles) in addition to the detection of filaments of the cosmic web.

\section{Conclusions and summary}
\label{conclusions}
This paper presented the first published general-purpose catalogues of filaments derived in the SDSS with the \disperse method, used in the works \citet{Malavasi2019,Tanimura2019disperse,Bonjean2019Filaments}, and how they have been derived. We provided a full investigation of the effect of the parameter choice for the \disperse algorithm on the skeleton reconstruction. We also provided a complete characterisation of the systematic problems that may affect our filaments, such as FoG effect redshift distortions and problems at the survey boundary. We characterised and validated the catalogue, ensuring the absence of strong problems in the filament samples used for the above mentioned works. We provide information on how the data may be accessed at \url{https://byopic.eu}.

With increased interest in the study of the cosmic web and with several wide-area galaxy surveys foreseen in the future (e.g. Euclid, \citealt{Laureijs2011Euclid}, Prime Focus Spectrograph, PFS, \citealt{Takada2014PFS}), the capability of creating catalogues for use of the community is vital to both pave the way for future studies and to improve the techniques for detecting the LSS. This paper provides several reference catalogues with which studies of the cosmic web can be performed and with which the development of new tools can be compared \citep[e.g.][]{Bonnaire2019}.

\begin{acknowledgements}
The authors thank the anonymous referee whose comments helped improve the quality of this work.

This research has been supported by the funding for the ByoPiC project from the European Research Council (ERC) under the European Union's Horizon 2020 research and innovation programme grant agreement ERC-2015-AdG 695561.

This work made use of the SZ-Cluster Database (\url{http://szcluster-db.ias.u-psud.fr}) operated by the Integrated Data and Operation Centre (IDOC) at the Institut de Astrophysique Spatiale (IAS) under contract with CNES and CNRS.

Funding for the SDSS and SDSS-II has been provided by the Alfred P. Sloan Foundation, the Participating Institutions, the National Science Foundation, the U.S. Department of Energy, the National Aeronautics and Space Administration, the Japanese Monbukagakusho, the Max Planck Society, and the Higher Education Funding Council for England. The SDSS Web Site is \url{http://www.sdss.org/}.

The SDSS is managed by the Astrophysical Research Consortium for the Participating Institutions. The Participating Institutions are the American Museum of Natural History, Astrophysical Institute Potsdam, University of Basel, University of Cambridge, Case Western Reserve University, University of Chicago, Drexel University, Fermilab, the Institute for Advanced Study, the Japan Participation Group, Johns Hopkins University, the Joint Institute for Nuclear Astrophysics, the Kavli Institute for Particle Astrophysics and Cosmology, the Korean Scientist Group, the Chinese Academy of Sciences (LAMOST), Los Alamos National Laboratory, the Max-Planck-Institute for Astronomy (MPIA), the Max-Planck-Institute for Astrophysics (MPA), New Mexico State University, Ohio State University, University of Pittsburgh, University of Portsmouth, Princeton University, the United States Naval Observatory, and the University of Washington.

Funding for SDSS-III has been provided by the Alfred P. Sloan Foundation, the Participating Institutions, the National Science Foundation, and the U.S. Department of Energy Office of Science. The SDSS-III web site is \url{http://www.sdss3.org/}.

SDSS-III is managed by the Astrophysical Research Consortium for the Participating Institutions of the SDSS-III Collaboration including the University of Arizona, the Brazilian Participation Group, Brookhaven National Laboratory, Carnegie Mellon University, University of Florida, the French Participation Group, the German Participation Group, Harvard University, the Instituto de Astrofisica de Canarias, the Michigan State/Notre Dame/JINA Participation Group, Johns Hopkins University, Lawrence Berkeley National Laboratory, Max Planck Institute for Astrophysics, Max Planck Institute for Extraterrestrial Physics, New Mexico State University, New York University, Ohio State University, Pennsylvania State University, University of Portsmouth, Princeton University, the Spanish Participation Group, University of Tokyo, University of Utah, Vanderbilt University, University of Virginia, University of Washington, and Yale University.

\end{acknowledgements}

\bibliographystyle{aa}
\bibliography{catalogue}

\appendix
\section{Filament distribution with the cleaned skeleton}
\label{cfig}
This appendix contains alternative versions of Figs. \ref{Fil_map_radec_legacy}, \ref{Fil_map_radec_lowzcmass}, \ref{Fil_map_raz_legacy}, and \ref{Fil_map_raz_lowzcmass} obtained with a cleaned skeleton to show an example of the distribution of secure filaments. The skeleton was cleaned from critical points (including bifurcations) and filaments affected by edges (see Section \ref{method_edgeeffects}) and minor problems (see Section \ref{method_minorissues}). The cleaned skeleton clearly lacks the long filaments that extend at the edge of the survey in regions that appear to be devoid of galaxies and other similar defects. The clear lack of filaments at low redshift is likely due to our conservative approach to eliminating edge-affected bifurcations. As mentioned in Section \ref{method_edgeeffects}, we remark that the fixed threshold of 100 Mpc (200 Mpc) for the Legacy MGS (LOWZ+CMASS) sample we adopted to clean bifurcations may remove too many of them, especially at low redshift, and a redshift-evolving criterion to remove bifurcations may be more indicated.

\begin{figure*}
\centering
\includegraphics[width = \linewidth, trim = 0cm 2cm 1cm 2cm, clip = true]{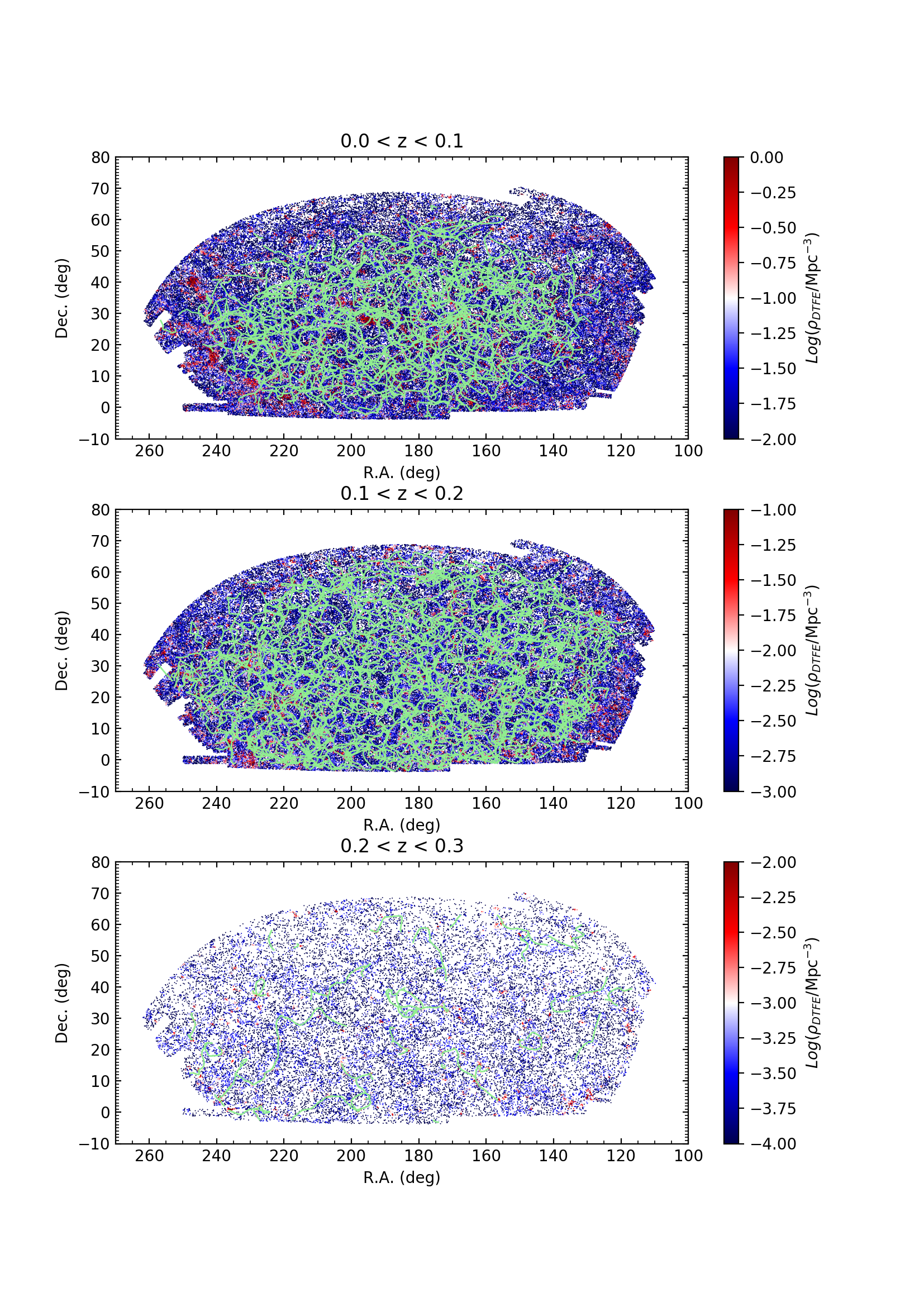}
\caption{Angular distribution of the filaments on the plane of the sky. The galaxy distribution is colour-coded according to the local density as measured by the DTFE, while filaments are over-plotted in green. Three ranges of redshift are considered for clarity, as marked on top of each panel. Only the Legacy MGS is considered. As an example, filaments detected with a $3\sigma$ persistence threshold and one smoothing cycle of the density field have been reported. The skeleton has been cleaned for edge effects and minor issues.}
\label{Fil_map_radec_legacy_cleaned}
\end{figure*}

\begin{figure*}
\centering
\includegraphics[width = \linewidth, trim = 0cm 2cm 1cm 2cm, clip = true]{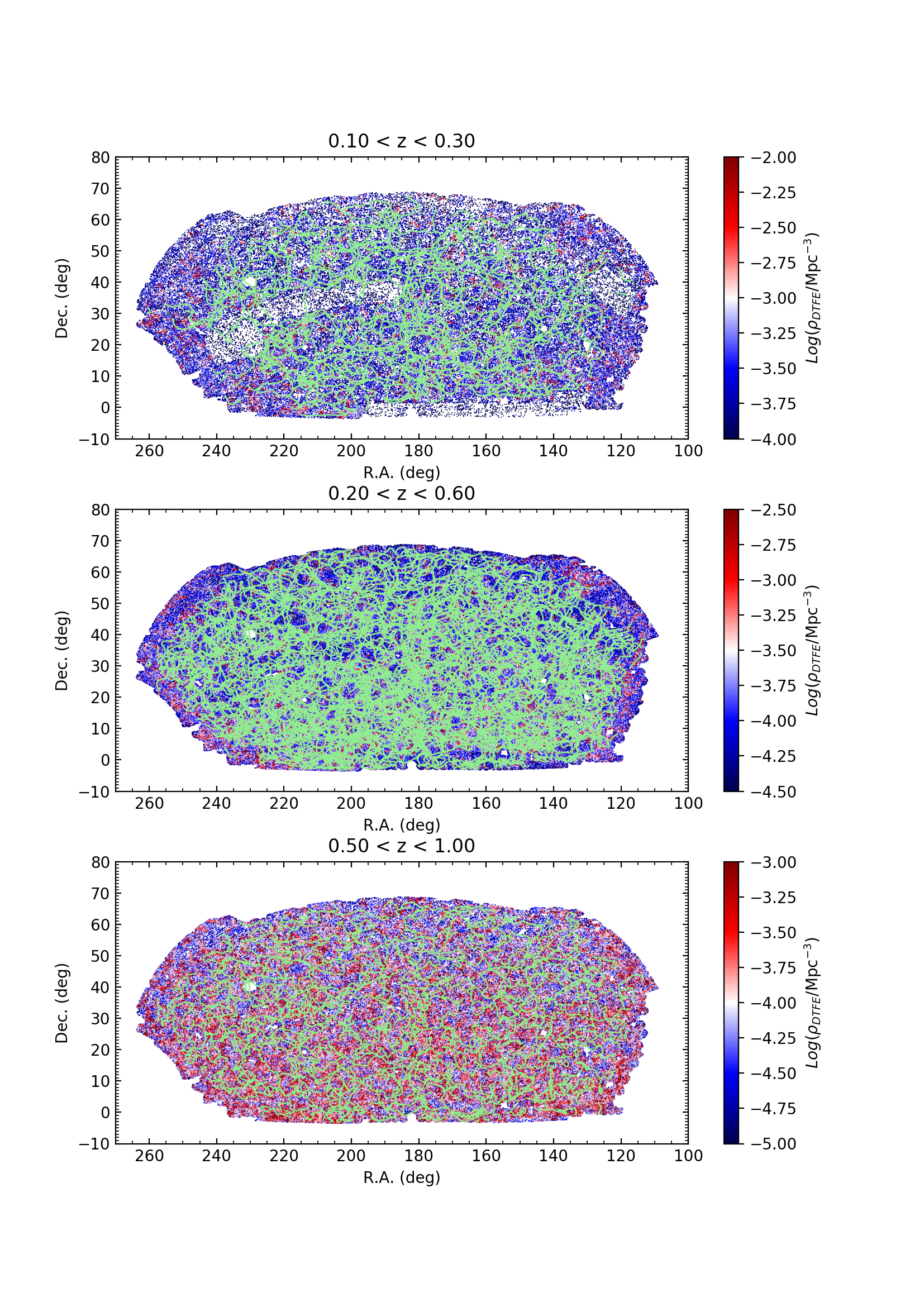}
\caption{Angular distribution of the filaments on the plane of the sky. The galaxy distribution is colour-coded according to their local density as measured by the DTFE, and filaments are over-plotted in green. Three ranges of redshift are considered for clarity, as marked at the top of each panel. Only the LOWZ+CMASS sample is considered. As an example, filaments detected with a $3\sigma$ persistence threshold and no smoothing of the density field have been reported. The skeleton has been cleaned from edge effects and minor problems.}
\label{Fil_map_radec_lowzcmass_cleaned}
\end{figure*}

\begin{figure*}
\centering
\includegraphics[width = \linewidth]{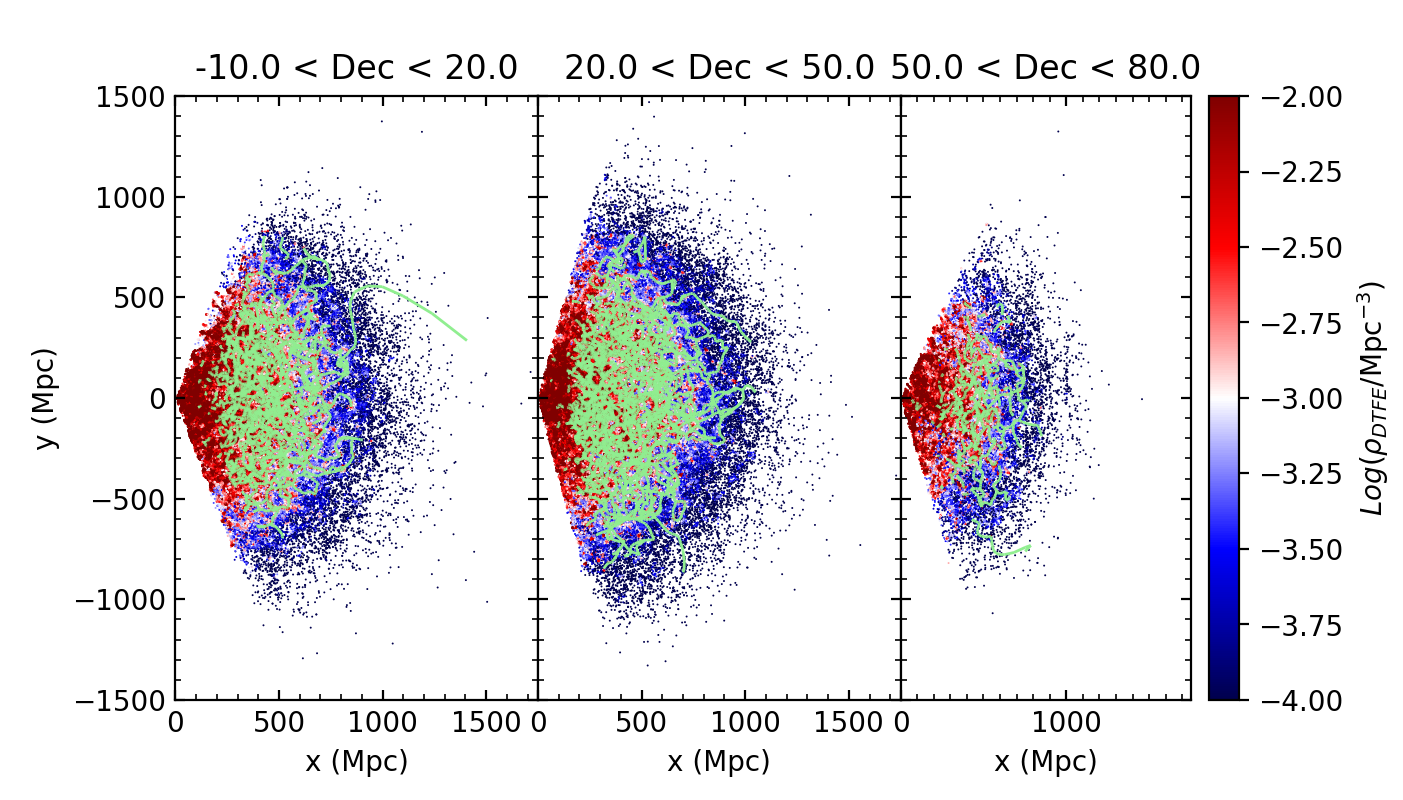}
\caption{Maps of filaments on a Cartesian $x-y$ plane. The $x$-axis is aligned with increasing redshift and the $y$-axis with increasing right ascension. The galaxy distribution is colour-coded according to local density as measured by the DTFE, and filaments are over-plotted in green. Three ranges of declination are considered for clarity, as marked at the top of each panel. Only the Legacy MGS is considered. As an example, filaments detected with a $3\sigma$ persistence threshold and one smoothing cycle of the density field have been reported. The skeleton has been cleaned from edge effects and minor problems.}
\label{Fil_map_raz_legacy_cleaned}
\end{figure*}

\begin{figure*}
\centering
\includegraphics[width = \linewidth]{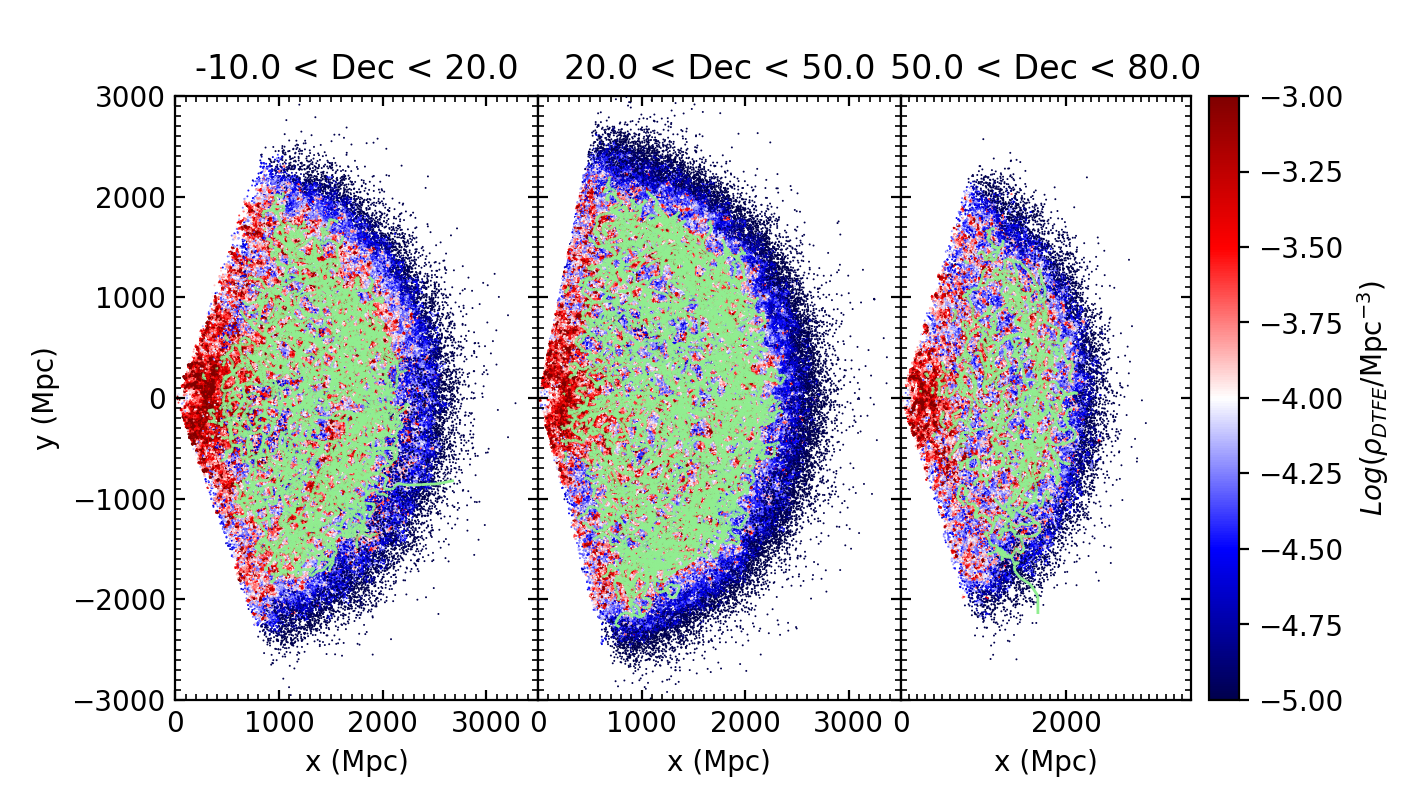}
\caption{Maps of filaments on a Cartesian $x-y$ plane. The $x$-axis is aligned with increasing redshift and the $y$-axis with increasing right ascension. The galaxy distribution is colour-coded according to local density as measured by the DTFE, and filaments are over-plotted in green. Three ranges of declination are considered for clarity, as marked at the top of each panel. Only the LOWZ+CMASS is considered. As an example, filaments detected with a $3\sigma$ persistence threshold and no smoothing of the density field have been reported. The skeleton has been cleaned from edge effects and minor problems.}
\label{Fil_map_raz_lowzcmass_cleaned}
\end{figure*}

\end{document}